\documentclass[twocolumn]{aastex62}

\usepackage{amsmath}

\graphicspath{{./}{figures/}}

\shorttitle{Observing SN Neutrinos}
\shortauthors{Suwa et al.}

\begin{document}

\title{Observing Supernova Neutrino Light Curves with Super-Kamiokande: Expected Event Number over 10 s}

\correspondingauthor{Yudai Suwa}
\email{suwa@yukawa.kyoto-u.ac.jp}

\author[0000-0002-7443-2215]{Yudai Suwa}
\affiliation{Department of Astrophysics and Atmospheric Sciences, Kyoto Sangyo University, Kyoto 603-8555, Japan}
\affiliation{Center for Gravitational Physics, Yukawa Institute for Theoretical Physics, Kyoto University, Kyoto, 606-8502 Japan}

\author [0000-0002-9224-9449]{Kohsuke Sumiyoshi}
\affiliation{National Institute of Technology, Numazu College of Technology, Shizuoka 410-8501, Japan}

\author[0000-0001-6330-1685]{Ken'ichiro Nakazato}
\affiliation{Faculty of Arts and Science, Kyushu University, Fukuoka 819-0395, Japan}

\author{Yasufumi Takahira}
\affiliation{Department of Physics, Okayama University, Okayama 700-8530, Japan}

\author[0000-0003-0437-8505]{Yusuke Koshio}
\affil{Department of Physics, Okayama University, Okayama 700-8530, Japan}
\affiliation{Kavli Institute for the Physics and Mathematics of the Universe (Kavli IPMU, WPI), Todai Institutes for Advanced Study, the University of Tokyo, Kashiwa 277-8583, Japan}

\author{Masamitsu Mori}
\affiliation{Department of Physics, Kyoto University, Kyoto 606-8502, Japan}

\author{Roger A. Wendell}
\affiliation{Department of Physics, Kyoto University, Kyoto 606-8502, Japan}
\affiliation{Kavli Institute for the Physics and Mathematics of the Universe (Kavli IPMU, WPI), Todai Institutes for Advanced Study, the University of Tokyo, Kashiwa 277-8583, Japan}



\begin{abstract}

Supernova neutrinos are crucially important to probe the final phases of massive star evolution. As is well known from observations of SN 1987A, neutrinos provide information on the physical conditions responsible for neutron star formation and on the supernova explosion mechanism. However,  there is still no complete understanding of the long-term evolution of neutrino emission in supernova explosions, although there are a number of modern simulations of neutrino radiation hydrodynamics, which study neutrino emission at times less than one second after the bounce. In the present work we systematically calculate the number of neutrinos that can be observed in Super-Kamiokande over periods longer than ten seconds using the database of \cite{naka13} anticipating that neutrinos from a Galactic supernova can be detected for several tens of seconds. We find that for a supernova at a distance of 10 kpc, neutrinos remain observable for longer than 30 s for a low-mass neutron star ($1.20M_\odot$ gravitational mass) and even longer than 100 s for a high-mass neutron star ($2.05M_\odot$). These scenarios are much longer than the observations of SN 1987A and longer than the duration of existing  numerical simulations. We propose a new analysis method based on the cumulative neutrino event distribution as a function of reverse time from the last observed event, as a useful probe of the neutron star mass. Our result demonstrates the importance of complete modeling of neutrino light curves in order to extract physical quantities essential for understanding supernova explosion mechanisms, such as the mass and radius of the resulting neutron star.

\end{abstract}



\section{Introduction} \label{sec:intro}

Neutrino bursts from core-collapse supernovae carry precious information about the central objects formed in these explosive phenomena \citep{kot06,mir16,hor18b}.  In the same way that optical light curves determine the type of the supernovae and allow for the study of the progenitor and nucleosynthesis, neutrino light curves contain important keys to solving open issues concerning the supernova explosion mechanism.  Since neutrinos play an essential role in supernova dynamics through trapping and emission processes in the central core,  escaping neutrinos retain the full information on what is happening deep inside \citep{jan17bk1,jan17bk2}.  These neutrinos are mainly emitted from the surface of the nascent neutron star and can be used to probe the thermal condition of the dense matter involved in the explosion mechanism.  Therefore, the detection of supernova neutrinos is a prime target of  neutrino astronomy \citep{kos92,sch12}.  

The observation of supernova neutrinos from SN 1987A \citep{hir87,bio87} established that it is possible to extract supernova physics from the emitted neutrinos. Indeed, timing and energy information from a handful of neutrino events has been used to establish the general core-collapse supernovae scenario \citep{sat87a,sat87b,bur87,blu88,jan89}. The duration of the neutrino burst over $\sim$10 s indicates the time scale of neutrino diffusion at high densities. The total energy of (3--6)$\times 10^{52}$ erg carried by the $\bar\nu_e$ flux, which is roughly 1/6 of the total neutrino flux, suggests the formation of a typical neutron star with gravitational binding energy of (2--3)$\times 10^{53}$ erg. 
In addition, the energy distribution of neutrinos detected in the range of 10--40 MeV demonstrates that a hot compact object evolves with a temperature of 3--5 MeV at the neutrino emitting region. Despite these successes, due to the limited size of the detectors at the time and the distance to SN 1987A, the total number of observed events is small and it is therefore not possible to draw a detailed picture of the explosion mechanism.

The next detection of supernova neutrinos will provide an enormous amount of information on the explosion and modern neutrino detectors are preparing to observe the time profile of the burst (the neutrino emission light curve). 
Super-Kamiokande will record the energies and directions of $\sim$10$^4$ neutrino events for a Galactic supernova~\citep{ike07}.  Loading  Gd in the detector will further enhance its ability to distinguish different species and improve pointing back to the collapsed star \citep{bea04}.\footnote{Multiple detectors allow us to perform triangulation with neutrinos, which potentially provides better spatial resolution \citep{brda18}.}  A supernova similar to SN 1987A in the Large Magellanic Cloud will produce $\sim$400 neutrino events in the detector, which is enough to observe the time profile of the neutrino burst. Such an observation will allow detailed exploration of the physics behind the supernova explosion.  

This exceptional opportunity necessitates systematic preparation of neutrino light curves in advance. That is, in order to extract the details of the supernova mechanism from observational data, systematic coverage of the impact of physical properties on the neutrino spectrum, such as the details of the progenitor and the resulting compact object, is essential.  
Providing a complete set of event rate predictions for both successful and failed supernova explosions from various progenitors is therefore desirable.
It is equally important to provide systematic predictions covering variations in the microphysics to extract information about state of the dense matter at the supernova's core.  

Although such a comprehensive study of the neutrino burst has been pursued for decades, the uncertainty surrounding the as-yet unknown explosion mechanism is the main obstacle to precise predictions of the neutrino signal.  Furthermore, correlated effects of physical parameters obscure the differences in the signatures of neutrino bursts in some situations.  At the time of SN 1987A, only a handful of spherically symmetric (1D) supernova simulations were available to infer the progenitor \citep{sat87a} and a series of proto-neutron star (PNS) models were used to put constraints on the central object \citep{bur88}.  Numerical studies in later years typically covered separate parts of the neutrino burst time profile: the early stage after the core bounce and the cooling of the PNS.  

Since shock propagation is the main issue in the supernova mechanism, most numerical studies follow its time evolution up to one second after the core bounce.  Accordingly, studies of neutrino emission have mostly been made at times around the bounce and neutronization burst \citep{myr90,tho03}.  The dependence of neutrino emission on the progenitor and the equation of state (EOS) has also been studied for short periods after the bounce \citep{tho03,sum05}.\footnote{See also \cite{odr04} and \cite{kat17} for the pre-collapse phase.} 
Under typical conditions such spherically symmetric studies have failed to produce an explosion and hence attention has been paid to only the early phases of collapse.

With the recent revelation of explosions driven by neutrino heating in two and three dimensions \citep{kot12,bur13,jan16}, multi-dimensional features of the neutrino burst have been explored to probe hydrodynamic instabilities such as the standing-accretion-shock instability (SASI), convection, sound wave reflection, and rotation \citep{mar09,lun12,suwa13,tam13,yok15,mir16,kur17,tak18}.  Note that most of these state-of-the-art simulations have been done for only limited times due to computational restrictions.  
As a result, despite the general importance of understanding neutrino emission from nascent compact objects, such studies have been confined to the period earlier than about one second after collapse. 

Studies of the long-term neutrino emission from the cooling of the PNS are important because the majority of supernova neutrinos come from this phase \citep{bur86,suz94,pon99}.  Although the neutrino luminosity decays rapidly with time, neutrinos produced at times between 1--20 s dominate the expected signal as was demonstrated by observations of SN 1987A.  Models of neutrino emission from PNS cooling have been used to assess the compact object at the center of the supernova by comparisons to data from SN 1987A \citep{bur87,bur88}. Extensive studies of neutrino emission have been made to probe the properties of dense matter \citep{sum95,rob12a,cam17,nak18} including hyperons and quarks \citep{pon01a,pon01b}, the influence of neutrino interactions~\citep{suz93,pin12,fis16b}, as well as the effects of convection \citep{rob12b}.\footnote{See \cite{kei95}, \cite{bau96b}, and \cite{sum07} for cases in which a black hole is formed.} 
A systematic study of various aspects of supernova neutrinos from early to late times based on sophisticated simulations is presented in~\cite{mir16}.
In these studies the initial conditions have been prepared using other supernova models, thereby separating the PNS cooling from the explosion mechanism. 

In order to explore the connection between progenitors and compact objects using the neutrino signal, it is essential to study the long-term neutrino emission starting from the initial stellar model and include the thermal evolution of the PNS born in the resulting explosion. 
Evolution from the supernova explosion to the formation of a neutron star has been studied in 1D \citep{tot98,hue10,fis10} for selected progenitors and to a limited extent in 2D \citep{suw14}.\footnote{See \cite{nak15} and \cite{hor18a} for limitations of these models.}  Neutrino light curves in the 1D models by \cite{tot98} and \cite{dal99} have been routinely used to evaluate the total event number at neutrino detectors \citep[see also][]{mir16}.  
Due to limitations in the ability of these predictions and their focus on explosion dynamics, the expected number of neutrino events at underground neutrino detectors such as IceCube \citep{tam13}, Super-Kamiokande \citep{ike07} and Hyper-Kamiokande \citep{hypk11,hypk16} has been estimated only for the dynamical phase using various models and numerical simulations to discuss these detectors' ability to probe the physics of the supernova. 
As a result, the long-term behavior of the neutrino burst from the PNS has not been studied in detail with modern simulations covering a variety of progenitors.  

Our first aim in the current study is, therefore, to predict the basic features of the expected number of events at Super-Kamiokande for the full time sequence of supernova models.  
We utilize neutrino emission properties from a supernova neutrino database \citep{naka13}, which covers the complete dynamical evolution of the event from gravitational collapse to the cooling of the PNS.  Through comparisons of event numbers for a set of progenitors systematically obtained from neutrino-radiation hydrodynamics (RHD) and PNS cooling simulations, we explore the differences in neutrino signals among models based on different progenitors. 
We discuss the time profile of the event rate for both early times around the bounce and the later phases of PNSs.  We estimate the basic features of the event rate evolution for various models to assess whether one can extract information on the progenitor from future observations of a supernova neutrino burst. This is an important step in discussing the possibility of studying the progenitor's mass, metallicity, and explosion timing, for instance, immediately after detection at Super-Kamiokande.  

Our second aim is to demonstrate the importance of the late phase of supernova neutrino emission.  We study the long-term evolution of PNSs over 100 s to evaluate the final phases of neutrino detection at Super-Kamiokande. In addition, new simulations of PNS cooling adopting different initial conditions have been performed to investigate weakening of the neutrino signal.  
We determine the timing of the last detected event to assess the time duration of neutrino emission and propose a backward-time analysis,  characterized by the cumulative event distribution as integrated backward in time from the last event, in order to disentangle the PNS properties.  
This approach is advantageous because the late phase cooling through diffusion is expected to be more quasi-static and simple \citep[see][for convection]{rob12a} than the early phase, where the hydrodynamic behavior of shock dynamics just after the bounce is highly complicated.  
Furthermore, this method may provide a basis for exploring dynamical situations, such as SASI and convection, in the early phase by extrapolation from the much simpler late-time neutrino light curves.

This paper is arranged as follows.  We describe the modeling of supernova neutrino light curves from core-collapse supernovae and PNS cooling in \S \ref{sec:simulation}.  In addition to the supernova neutrino database in \S \ref{sec:SNdatabase}, we explain additional modeling used to extract properties of the PNS with various masses in \S \ref{sec:PNSCdata}. 
Neutrino detection at Super-Kamiokande is discussed in \S \ref{sec:SK}. 
In \S \ref{sec:event_number}, we provide basic information concerning the expected event rates for a set of supernovae taken from the neutrino database used in conjunction with additional models.  
We also discuss the differences and similarities among various models and evaluate the feasibility of distinguishing model parameters with the neutrino light curve.
Furthermore, we examine the long-term phase of recent PNS cooling models to assess the neutrino emission's dependence on mass. 
In \S \ref{sec:howlong}, we study the final phase of neutrino detection at Super-Kamiokande and discuss the timing of the last event as a function of the distance to the supernova in \S \ref{sec:fadeout}.  This analysis illustrates the importance of studying the long duration of galactic events.  
In \S \ref{sec:backward} we propose a backward-time analysis of the neutrino signal to explore its dependence on the PNS model.  
We evaluate the cumulative event distribution by integrating the number of observed events backward in time from the last detection. 
In this way the cumulative event history is used to discriminate different PNS models.  
Finally, in \S \ref{sec:analysis} we describe our data analysis strategy for Galactic supernova bursts before summarizing in \S \ref{sec:summary}.  

\section{Neutrino emission simulations} \label{sec:simulation}
In this paper, we utilize the supernova neutrino emission provided by the database briefed in \S~\ref{sec:SNdatabase}. 
We also consider the long-term ($\ge$20 s) development of the neutrino emission utilizing numerical data from the thermal evolution of quasi-static PNSs as described in \S~\ref{sec:PNSCdata}.

\subsection{Supernova neutrino database} \label{sec:SNdatabase}

In the supernova neutrino database \citep{naka13}, neutrino spectra for various scenarios are provided as a function of time up until 20 s after the bounce. For this purpose, numerical simulations of the neutrino-RHD of stellar cores and the thermal evolution of quasi-static PNSs with neutrino emission are combined. 
In both simulations, the EOS from \citet{she98a,she98b} is utilized.\footnote{Note that from observations of GW170817 EOSs with relatively large neutron star radii, including the EOS used in this study, are strongly constrained. However, in order to use \cite{naka13} we utilize Shen's EOS. The dependence of the neutrino spectrum on the choice of EOS will be presented in a forthcoming paper.} 
Progenitor models were made from four progenitor masses ($M_{\rm ZAMS}=13$, 20, 30, and $50M_\odot$) and two metallicities ($Z= 0.02$ or 0.004) adopted from numerical results from the stellar evolution code \citep{umeda12}. 
While the $30M_\odot$ and $Z=0.004$ progenitor is a black hole-forming model due to its large iron-core mass, the other seven progenitors are models for ordinary core-collapse supernovae. 
In this paper, we utilize only models with $Z=0.02$ (solar metallicity).

In constructing the supernova neutrino database, neutrino emission in the early phase is computed with an implicit Lagrangian code for general relativistic neutrino-RHD, which solves the neutrino Boltzmann equations and the dynamics of spherical gravitational collapse simultaneously \citep{yama97,yama99,sum05}. 
This code follows the neutrino distribution functions for four species,  $\nu_e$, $\bar\nu_e$, $\nu_x$ ($=\nu_\mu$, $\nu_\tau$), and $\bar\nu_x$ ($=\bar\nu_\mu$, $\bar\nu_\tau$), over a discrete grid in energy and angle to solve the neutrino Boltzmann equations. 
The difference between $\nu_x$ and $\bar\nu_x$ is minor and 
accordingly they are treated collectively and denoted as $\nu_x$ in the supernova neutrino database. 
The following neutrino reactions are considered: (1) electron-type neutrino absorption on neutrons and its inverse, (2) electron-type anti-neutrino absorption on protons and its inverse, (3) neutrino scattering on nucleons, (4) neutrino scattering on electrons, (5) electron-type neutrino absorption on nuclei, (6) neutrino coherent scattering on nuclei, (7) electron-positron pair annihilation and creation, (8) plasmon decay and creation, and (9) neutrino bremsstrahlung. 
Further details are presented in \cite{sum05}.

In contrast to the early phase described above, neutrino emission during the late phase is computed using the general relativistic quasi-static evolutionary code of neutrino diffusion \citep{suz93,suz94,suz05}. 
In order to follow the quasi-static evolution of PNSs, this code solves the hydrostatic structure of the PNS using the Oppenheimer-Volkoff equation at each time step taking into account deleptonization and entropy evolution from neutrino transfer with a Henyey-type method.
The neutrino transfer utilizes a multi-group flux-limited diffusion scheme assuming spherical symmetry in general relativity and adopts the flux limiter in \cite{may87}. 
In this method, the Boltzmann equations in their angle-integrated form are treated taking into account the energy dependence of $\nu_e$, $\bar\nu_e$, and $\nu_x$, whereas $\nu_\mu$, $\nu_\tau$, $\bar\nu_\mu$, and $\bar\nu_\tau$ are treated collectively as $\nu_x$. The same neutrino reactions as in the  general relativistic neutrino-RHD code above are included in the general relativistic quasi-static evolutionary code.

\begin{figure*}[htbp]
\centering
\includegraphics[width=.6\textwidth]{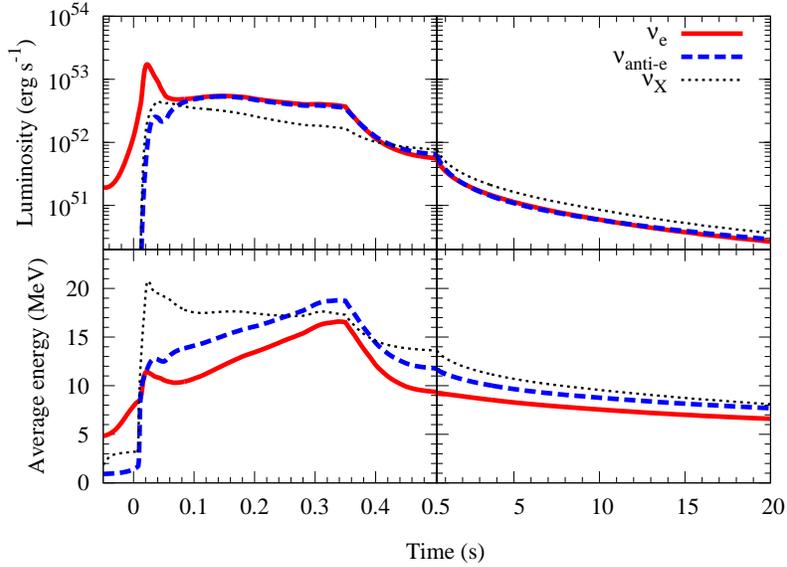}
\caption{Neutrino luminosities (top panels) and average energies (bottom panels) as a function of time after bounce 
for the 13$M_{\odot}$, $Z=0.02$, $t_{\rm revive}=300$~ms model.}
\label{fig:1303}
\end{figure*}

In addition to the computations described above, neutrino emission in the intermediate regime is evaluated by interpolating between the two phases. Since the neutrino-RHD simulations for the early phase are performed under the assumption of spherical symmetry, the accretion rate, which is converted to neutrino luminosity, will be overestimated. 
Mass accretion will be reduced due to multi-dimensional effects, such as  convection and SASI. On the other hand, the neutrino emission due to the matter fallback is not included in the PNS cooling simulations for the late phase. Therefore, the neutrino emissions obtained by the two simulations can be regarded as upper and lower limits. 
While the neutrino-RHD simulations account for the neutrino emission before  shock revival, the neutrino light curves from the PNS cooling simulations are reasonable for times after the shock revival. On the basis of these considerations, the neutrino light curves of the early and late phases are interpolated by an exponential function assuming  shock  revival at either $t_{\rm revive}=100$, 200, or 300 ms after bounce. In Figure \ref{fig:1303}, a typical neutrino light curve obtained by this procedure is displayed.

\subsection{Proto-neutron star cooling} \label{sec:PNSCdata}

In order to investigate the long-term (over 100 s) behavior of PNS cooling, we utilize the numerical code in \cite{nak18}.
This part of the neutrino data corresponds to the late phase of \cite{naka13}. 
In this model, the result of the general relativistic neutrino-RHD simulation obtained with the numerical code described in \S~\ref{sec:SNdatabase} is used as an initial condition. With the EOS from \cite{she11}, core-collapse of the progenitor with $15M_\odot$ from \cite{ww95} is followed until $t=0.3$~s, as measured from the bounce. 
Then, the entropy and electron fraction profiles for the central region inside the shock wave are adopted as initial conditions for the PNS cooling simulation. Since the shock wave is stalled at the baryon mass coordinate of $m_b=1.47M_\odot$ at $t=0.3$~s, we consider a PNS with $M_b=1.47M_\odot$,  corresponding to a gravitational mass of $1.35M_\odot$, in the following. This model is denoted as 147S hereafter and it is also described in \cite{nak18}.

\begin{figure}[htbp]
\includegraphics[width=0.45\textwidth]{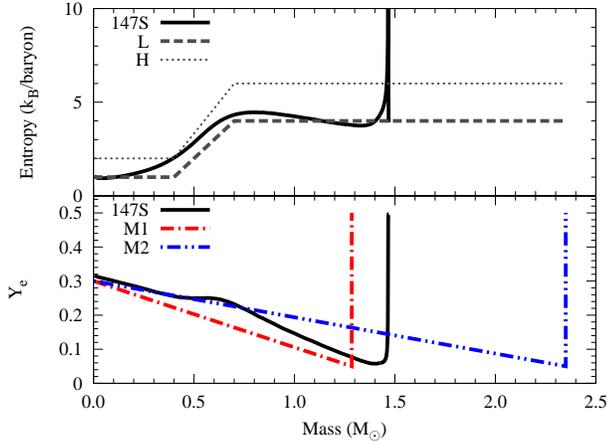}
\caption{Entropy (upper) and electron fraction (lower) profiles as a function of the baryonic mass coordinate $m_b$. In both panels, thick solid lines are for the model in \cite{nak18} with the EOS from \cite{she11}. In the upper panel, thin dotted and thick dashed lines correspond to models with $(s_1,s_2)=(2k_B, 6k_B)$ (H) and $(1k_B, 4k_B)$ (L), respectively. In the lower panel,  the red dot-dashed and the blue dot-dot-dashed lines correspond to models with $M_b=1.29M_\odot$ (M1) and $M_b=2.35M_\odot$ (M2), respectively.}
\label{fig:initprof}
\end{figure}

In addition to the models above, we also consider PNS models with baryon masses of $M_b=1.29M_\odot$ and $2.35M_\odot$, which correspond to gravitational masses of $1.20M_\odot$ and $2.05M_\odot$, respectively.
Note that the chosen mass range is based on recent observations of high-mass and low-mass pulsars in binary systems.
The highest mass is $\approx 2.0M_\odot$ \citep{demo10,antoni13}\footnote{Recently, a massive NS with $2.17^{+0.11}_{-0.1}M_\odot$ \citep{tha19} has been discovered.} and the lowest mass is $\approx 1.2M_\odot$ \citep{marti15}.\footnote{Theoretical estimations of the minimum mass of a neutron star are consistent with observations \citep{suwa18}.}
To construct these models, we perform new simulations in the same way as in \cite{nak18} adopting the initial entropy and electron fraction profiles given by
\begin{subequations}
\begin{align}
 & s(m_b) = \begin{cases}
    s_1 \quad  (0 \leq m_b \leq 0.4M_\odot) \\
    \displaystyle\frac{s_1(0.7M_\odot-m_b)+s_2(m_b-0.4M_\odot)}{0.3M_\odot} \\
    \qquad (0.4M_\odot \leq m_b \leq 0.7M_\odot) \\
    s_2 \quad (0.7M_\odot \leq m_b \leq M_b)
  \end{cases},\\
 & Y_e(m_b) = \frac{0.3(M_b-m_b)+0.05m_b}{M_b},
\end{align}
\label{eq:initprof}
\end{subequations}
where $s(m_b)$ and $Y_e(m_b)$ are the entropy per baryon and the electron fraction, respectively, at the baryon mass coordinate $m_b$. In this study, we consider two cases for the entropy; $(s_1,s_2)=(1k_B, 4k_B)$ and $(2k_B, 6k_B)$ are chosen as low-entropy and high-entropy cases, respectively. Here $k_B$ is the Boltzmann constant. 
In Figure~\ref{fig:initprof}, the profiles of Eq.~(\ref{eq:initprof}) are shown with the initial condition of PNS cooling in \cite{nak18}. For model names, we use MXY, in which X=1 and 2 denote $M_b=1.29M_\odot$ and $M_b=2.35M_\odot$, and Y=L,H denotes low- and high-entropy cases, respectively.

\begin{figure}[htbp]
\includegraphics[width=0.45\textwidth]{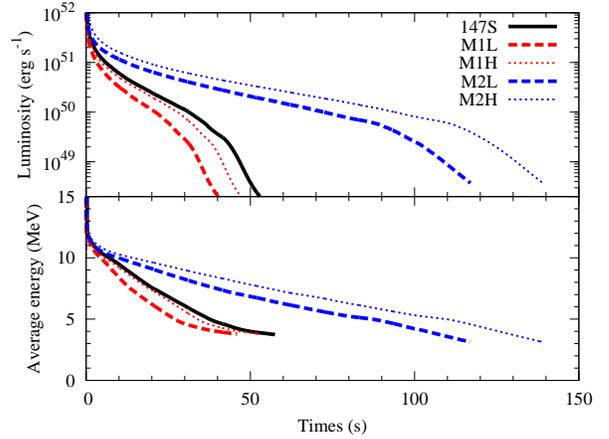}
\caption{Luminosity (upper) and average energy (lower) of $\bar\nu_e$ as a function of time from the birth of the PNS. Thick solid lines are for the model in \cite{nak18} with the EOS from \cite{she11}, red and blue lines are for models with initial conditions in Eq.~(\ref{eq:initprof}).
 Red lines are for low-mass ($M_b=1.29M_\odot$; M1) models and blue lines are for high-mass ($M_b=2.35M_\odot$; M2) models. The thick dashed lines are for the low-entropy (L) model and the thin dotted lines are for high-entropy (H) model.
}
\label{fig:nulc-pns}
\end{figure}

Figure~\ref{fig:nulc-pns} shows the $\bar\nu_e$ luminosity and average energy  evolution for the models described above. The average energy is calculated using the energy and number fluxes. 
Models with a PNS of $M_b=2.35M_\odot$ (blue lines) show longer neutrino emission than those with less massive,  $M_b=1.29M_\odot$, PNS (red lines).  Though the models with a higher initial entropy (indicated by thin dotted lines) imply longer emission, the impact is minor compared to the mass dependence. 
This indicates that the neutrino emission timescale contains information on the PNS, especially its mass.

\section{Detection at Super-Kamiokande}\label{sec:SK}

The Super-Kamiokande detector, which is located 1,000~m underground (2,700~m water equivalent) in the Kamioka mine in Gifu Prefucture, Japan, is a cylindrical tank (39.3~m in diameter and 41.4~m in height) filled with 50~kilotons of ultra-pure water~\citep{2003NIMPA.501..418F}.
The experiment started in 1996, and was shutdown for the latter half of 2018 for refurbishment ahead of a planned upgrade, known as SK-Gd, to load gadolinium in the detector's water. 
However, in this paper only simulations with pure water are performed.
The detector is divided into two regions called the inner and outer detectors, to distinguish real neutrino interactions from cosmic ray muon backgrounds.
The inner detector is lined with 11,129 20-inch photo multiplier tubes (PMT) and the outer detector uses 1,885 8-inch PMTs.
Cherenkov light generated by charged particles emerging from neutrino interactions in water is observed by the PMTs and used to reconstruct the neutrino signal.
The fiducial volume used in typical data analyses is 22.5~kilotons, defined as the volume more than 2~m from the inner wall of the inner detector, in order to ensure stable reconstruction performance and to reduce backgrounds from radioisotopes (RI).
However, for burst events like a supernova explosion, this kind of transient background will be negligible and therefore, the entire 32.5~kton volume of the inner detector is used in this paper.
The energy threshold for solar neutrino analysis in Super-Kamiokande is 4.0~MeV  (total electron energy), while 5~MeV is used in this paper to avoid RI background contamination completely.

There are several neutrino interactions in the relevant energy region of supernova neutrinos, a few to a few tens of MeV.
Inverse beta decay (IBD) with free protons, elastic scattering on electrons, and nuclear interactions with oxygen are typical examples.
The dominant signal is from IBD ($\bar{\nu_e} + p \rightarrow e^+ + n$) interactions, whose cross section has been calculated in \cite{1999PhRvD..60e3003V}.
Since the IBD cross section is about 10 times larger than other interactions, it is the only interaction channel considered in this paper. 
We note that the updated IBD cross section calculation in \cite{stru03} yields the same cross section as \cite{1999PhRvD..60e3003V} below 40 MeV, and would therefore not change the estimates in this paper.

The next-generation water Cherenkov detector, Hyper-Kamiokande, has been proposed in \citep{hypk16}.
Its detection principle is the same as Super-Kamiokande's, but with a  total inner detector volume of 220~kilotons.
Accordingly, we include sensitivity estimations for it in this paper as well.

\section{Expected event rates} \label{sec:event_number}

\begin{table*}[htbp]
    \centering
    \caption{Number of events for a supernova at 10kpc.}
    \begin{tabular}{lrrrrrrrrr}
    \hline
    Model & $M_{\rm ZAMS}$ & $t_{\rm revive}$ & $M_{\rm NS,g}$ & $N_{\rm tot}$ & $N(0\le t\le 0.3)$ & $N(0.3\le t\le 1)$ & $N(1\le t\le 10)$ & $N(10\le t\le 20)$ & $N(20\le t)$\\
     & ($M_\odot$) & (ms) & ($M_\odot$) \\
    \hline
   N13t100 &    13 &   100 &   1.39 &   3067.2 &  1210.5 (39.5\%) &   475.9 (15.5\%) &  1087.2 (35.4\%) &   293.6 ( 9.6\%) &     --- (  --- )  \\
   N13t200 &    13 &   200 &   1.46 &   3676.6 &  1672.8 (45.5\%) &   507.6 (13.8\%) &  1165.2 (31.7\%) &   331.1 ( 9.0\%) &     --- (  --- )  \\
   N13t300 &    13 &   300 &   1.50 &   4246.4 &  1807.2 (42.6\%) &   895.2 (21.1\%) &  1192.4 (28.1\%) &   351.7 ( 8.3\%) &     --- (  --- )  \\
   N20t100 &    20 &   100 &   1.36 &   2890.6 &  1089.7 (37.7\%) &   468.7 (16.2\%) &  1052.7 (36.4\%) &   279.4 ( 9.7\%) &     --- (  --- )  \\
   N20t200 &    20 &   200 &   1.42 &   3342.3 &  1437.8 (43.0\%) &   481.5 (14.4\%) &  1113.4 (33.3\%) &   309.6 ( 9.3\%) &     --- (  --- )  \\
   N20t300 &    20 &   300 &   1.45 &   3669.8 &  1525.7 (41.6\%) &   695.1 (18.9\%) &  1126.7 (30.7\%) &   322.4 ( 8.8\%) &     --- (  --- )  \\
   N30t100 &    30 &   100 &   1.49 &   3807.4 &  1649.9 (43.3\%) &   550.1 (14.4\%) &  1252.6 (32.9\%) &   354.8 ( 9.3\%) &     --- (  --- )  \\
   N30t200 &    30 &   200 &   1.66 &   5551.4 &  2952.4 (53.2\%) &   691.9 (12.5\%) &  1453.5 (26.2\%) &   453.6 ( 8.2\%) &     --- (  --- )  \\
   N30t300 &    30 &   300 &   1.78 &   7332.8 &  3363.4 (45.9\%) &  1919.6 (26.2\%) &  1533.4 (20.9\%) &   516.4 ( 7.0\%) &     --- (  --- )  \\
   N50t100 &    50 &   100 &   1.52 &   3788.9 &  1542.3 (40.7\%) &   553.2 (14.6\%) &  1314.8 (34.7\%) &   378.5 (10.0\%) &     --- (  --- )  \\
   N50t200 &    50 &   200 &   1.63 &   4883.1 &  2399.6 (49.1\%) &   616.1 (12.6\%) &  1428.4 (29.3\%) &   439.0 ( 9.0\%) &     --- (  --- )  \\
   N50t300 &    50 &   300 &   1.69 &   5952.3 &  2657.4 (44.6\%) &  1352.7 (22.7\%) &  1466.4 (24.6\%) &   475.9 ( 8.0\%) &     --- (  --- )  \\
      147S &   --- &   --- &   1.35 &   2205.4 &     --- (  --- ) &   434.3 (19.7\%) &  1278.5 (58.0\%) &   345.1 (15.6\%) &   147.5 ( 6.7\%)  \\
       M2H &   --- &   --- &   2.05 &   8032.8 &     --- (  --- ) &  1554.6 (19.4\%) &  2998.7 (37.3\%) &  1268.3 (15.8\%) &  2211.2 (27.5\%)  \\
       M1H &   --- &   --- &   1.20 &   2390.7 &     --- (  --- ) &   825.5 (34.5\%) &  1173.9 (49.1\%) &   288.0 (12.0\%) &   103.3 ( 4.3\%)  \\
       M2L &   --- &   --- &   2.05 &   4734.9 &     --- (  --- ) &   674.5 (14.2\%) &  2008.3 (42.4\%) &   867.1 (18.3\%) &  1185.0 (25.0\%)  \\
       M1L &   --- &   --- &   1.20 &   1382.8 &     --- (  --- ) &   376.5 (27.2\%) &   824.7 (59.6\%) &   148.4 (10.7\%) &    33.2 ( 2.4\%)  \\
    \hline
    \end{tabular}
    \tablecomments{
    $M_{\rm ZAMS}$ is the zero-age main sequence mass of the progenitor model.
    $t_{\rm revive}$ is the shock revival time.
    $M_{\rm NS,g}$ is the gravitational mass of PNS.
    These three numbers are taken from \cite{naka13}.
    $N_{\rm tot}$ is the total number of neutrinos. $N(t_{\rm min}\le t \le t_{\rm max})$ gives the number of events between $t_{\rm min}$ and $t_{\rm max}$, which are in seconds. Numbers in brackets are percentage of the total number. For models of the form N13t100 and similar, there is only data for $t< 20$s so no event rate is estimated at later times.
    Conversely, for models like M2H and similar, there are only calculations for the PNS cooling phase, so event rates before 0.3 s are not given.
    }
    \label{tab:model}
\end{table*}

\subsection{Results for the neutrino database}
\label{sec:eventno_database}

\begin{figure}[htbp]
\includegraphics[width=.45\textwidth]{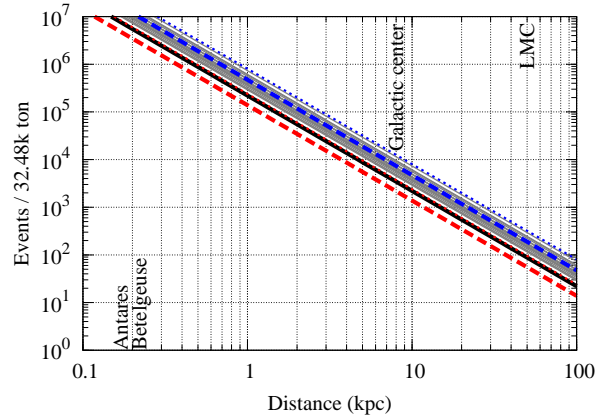}
\caption{Total expected number of IBD events as a function of the distance to the burst source for the models in the supernova neutrino database (gray lines). Colored lines correspond to the model calculations described in \S \ref{sec:PNSCdata}. Note that they do not include the early phase neutrinos, only late phase, so these models show systematically smaller event numbers than those in \cite{naka13} (represented by the gray band).
}
\label{fig:sn_total_event}
\end{figure}

We describe the features of the expected number of events for the series of models from the supernova neutrino database (see Table \ref{tab:model}). 
We select a set of models with a single metallicity ($Z=0.02$) and focus on four progenitor models (13, 20, 30, 50M$_{\odot}$) with freedom to choose the shock revival time.
This set covers a variety of density profiles of progenitor models and a range of PNS remnant masses. 
The density profile affects the luminosity through matter accretion right after the core bounce (early phase).  The remnant mass is determined by the progenitor model and shock revival time and affects the long-term behavior of the luminosity (late phase) via the total binding energy \citep{naka13}. 
In this section, the distance to supernova is set to 10 kpc, except for Figure \ref{fig:sn_total_event}.
Neutrino oscillations are not included, because they are not expected to significantly change the long-term evolution of neutrino light curves (see Sec. \ref{sec:backward}).
Detailed studies of the early-phase with the neutrino oscillation will be reported in a separate study. 

\begin{figure}[htbp]
\includegraphics[width=.45\textwidth]{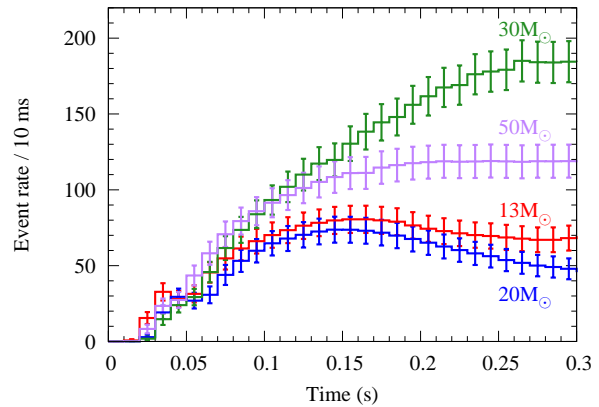} 
\caption{Expected number of IBD events as a function of time after bounce in the early phase for a supernova at 10 kpc in the 13, 20, 30, 50M$_{\odot}$ models are shown in the red, blue, green, and purple lines, respectively ($Z=0.02$, $t_{\rm revive}=300$~ms). The error bars are based on Poisson statistics.}
\label{fig:sn_early_event1}
\end{figure}

We show in Figure \ref{fig:sn_early_event1} the expected number of IBD events  (i.e. $\bar{\nu}_{e}$ interactions) as a function of time after the bounce.
In the early phase, up to 300 ms after the core bounce, the neutrino signal carries information on the core bounce and accretion onto it.  The rise of $\bar{\nu}_{e}$ interactions reflects components arising from thermal pair production and positron capture on protons in the accreting matter.  
We choose $t_{\rm revive}=300$ ms in this plot to examine the difference of accretion luminosities among the progenitors.  
The number of events rises quickly for 30M$_{\odot}$ and 50M$_{\odot}$ models as compared with those for 13M$_{\odot}$ and 20M$_{\odot}$, reflecting different rates of accretion.  
As discussed in the literature \citep{tho03,oco13,nak15,suwa16},  
the early phase event rate rise may probe the progenitor properties through the accretion luminosity, $L_{\rm acc}=GM\dot{M}/R$.

\begin{figure}[htbp]
\includegraphics[width=.45\textwidth]{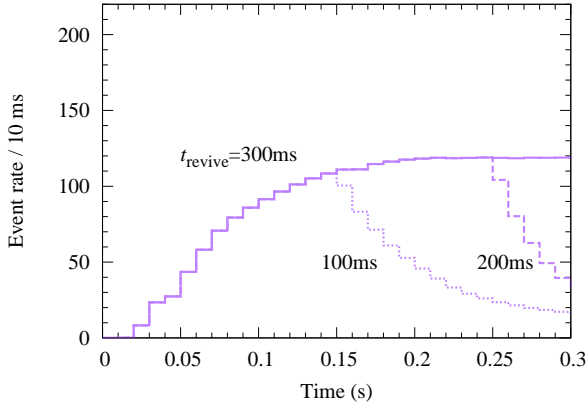} 
\caption{Same as Figure \ref{fig:sn_early_event1}, but  for the expected number of IBD events as a function of time after bounce in the early phase for a supernova at 10 kpc in the 50M$_{\odot}$ model ($Z=0.02$) for $t_{\rm revive}=100, 200, 300$~ms with dotted, dashed and solid line, respectively.}
\label{fig:sn_early_event2}
\end{figure}

The time when the event rate drops depends on the shock revival time, which is shown in Figure \ref{fig:sn_early_event2}.  
If the shock wave stalls around $t_{\rm revive}=300$~ms, the event rates stay at a certain level due to continued accretion. 
In the case of $t_{\rm revive}=100$~ms or 200 ms, the event rates rapidly decrease because accretion ends as the shock is revived in our model. 
The drop in the event rate is seen to correspond with the transition from the accretion phase to the diffusion phase.

We expect to detect such a luminosity transition (event rate transition) by observing the change in the neutrino light curve when the shock revives and accretion halts.  
Although the current database is based on 1D core-collapse dynamics and PNS cooling models, we envisage this transition exists even under more complicated situations as seen in modern 2D/3D simulations.  
We remark that one expects more variations in the event numbers in  2D and 3D simulations through hydrodynamic instabilities and non-uniform accretion with a deformed shock geometry \citep[e.g.,][]{tam13,tak18}. Our analysis here can be considered as a basis for studying such hydrodynamic complications by comparison of spherically symmetric and multi-dimensional simulations. 

\begin{figure}[htbp]
\includegraphics[width=.45\textwidth]{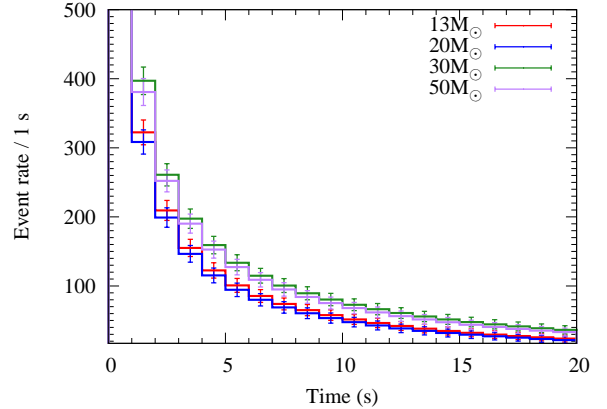} 
\caption{Expected number of IBD events as a function of time after bounce in the late phase for a supernova at 10 kpc in the 13, 20, 30, 50M$_{\odot}$ models shown by the red, blue, green, and purple lines, respectively ($Z=0.02$, $t_{\rm revive}=300$~ms).}
\label{fig:sn_late_event1}
\end{figure}

In the late phase (up to 20~s) of the evolution, the neutrino signal reflects the properties of the cooling PNS.  
A gradual decrease in the neutrino luminosity originates from the diffusion of neutrinos from the central part of the supernova. 
The luminosity depends mainly on the mass of PNS born in the collapse of the progenitor. 
In Figure \ref{fig:sn_late_event1}, the time profile of the expected number of events is shown for the progenitor models of 13--50M$_{\odot}$ with $t_{\rm revive}=300$~ms.  
The shape of the time profiles are similar among the four models, though their amplitudes depend on the PNS mass.  The number of events is largest for the 30M$_{\odot}$ model, which has a remnant neutron star with a gravitational mass of 1.78M$_{\odot}$, and is smallest for the 20M$_{\odot}$ model with a 1.45M$_{\odot}$ PNS.  

\begin{figure}[htbp]
\includegraphics[width=.45\textwidth]{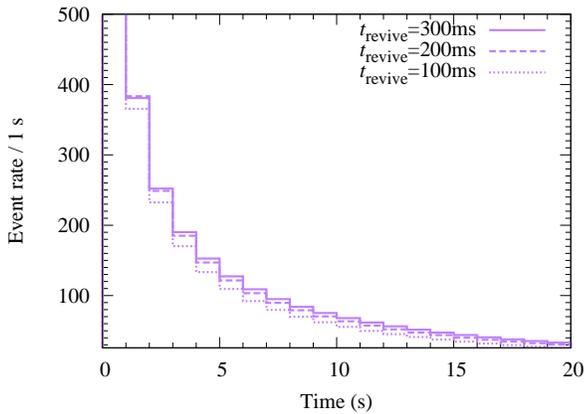} 
\caption{Expected number of IBD events as a function of time after bounce in the late phase for a supernova at 10 kpc in the 50M$_{\odot}$ model ($Z=0.02$) for $t_{\rm revive}=$100, 200, 300~ms shown by the dotted, dashed, and solid lines, respectively.}
\label{fig:sn_late_event2}
\end{figure}

The number of events also depends on the shock revival time, which determines the remnant mass by stopping accretion even within the same progenitor model.
In Figure \ref{fig:sn_late_event2}, we show how the expected number of events depends on the shock revival time for the 50M$_{\odot}$ model with three different PNS masses, (1.52M$_{\odot}$, 1.63M$_{\odot}$ and 1.69M$_{\odot}$, for $t_{\rm revive}=$100, 200, 300~ms, respectively. 
The largest PNS mass leads to the largest number of events because it represents the largest release of gravitational energy.  
Therefore, the late phase of the neutrino light curve of neutrinos is important to extract the properties of the compact object left after a supernova explosion.

Note that the properties of the remnant are in principle determined by the details of the explosion mechanism via the collapse and bounce of the progenitor.  
In this sense, the shock revival time is a simplified method for constructing a set of PNSs in 1D explosion models for study.  
In order to extract the properties of the remnant from an observation, one needs to carefully explore the impact of the remnant's parameters on the neutrino time profile. 
To study such variations in the profile, we explore the longer-time behavior in later sections.

In Figure \ref{fig:sn_total_event}, we show the  total expected number of IBD events in Super-K as a function of the distance to the supernova neutrino burst.  
The total number is obtained by the time integral of the event rates up to 20 s, the end time of the database.  
Each line corresponds to the total for one of the models in the supernova neutrino database.  
Typically $\sim4\times10^{3}$ events are expected for an event at a distance of 10 kpc.
However, the total ranges by a factor of 5 depending on the remnant mass of  the progenitor.  
Among the models,
the highest event rate is seen for the 30M$_{\odot}$ model with $t_{\rm revive}=300$ ms and the smallest is from the 20M$_{\odot}$ model with $t_{\rm revive}=100$ ms.  
The corresponding PNS masses range from 1.36M$_{\odot}$ to 1.78M$_{\odot}$ in the database.  

\subsection{Results for new PNS cooling models}
\label{sec:eventno_protoNS}

We further investigate the event rates of neutrino bursts using the PNS models in \S \ref{sec:PNSCdata} to study the details of late phase detection. 
In order to discuss the duration of the neutrino burst and to extract the properties of the remnant using the backward-time method (\S \ref{sec:backward}) we would like to determine the timing of the last detected event (Section \ref{sec:howlong}) .  
From the studies above using the neutrino database, the event rate is about 0.1 event/10 ms at 20 s in Figs. \ref{fig:sn_late_event1} and \ref{fig:sn_late_event2}.  
Hence, neutrino detection is expected to continue for even later times for a 10 kpc event and we need to explore time profiles over 50--100 s.  

\begin{figure}[htbp]
\includegraphics[width=.45\textwidth]{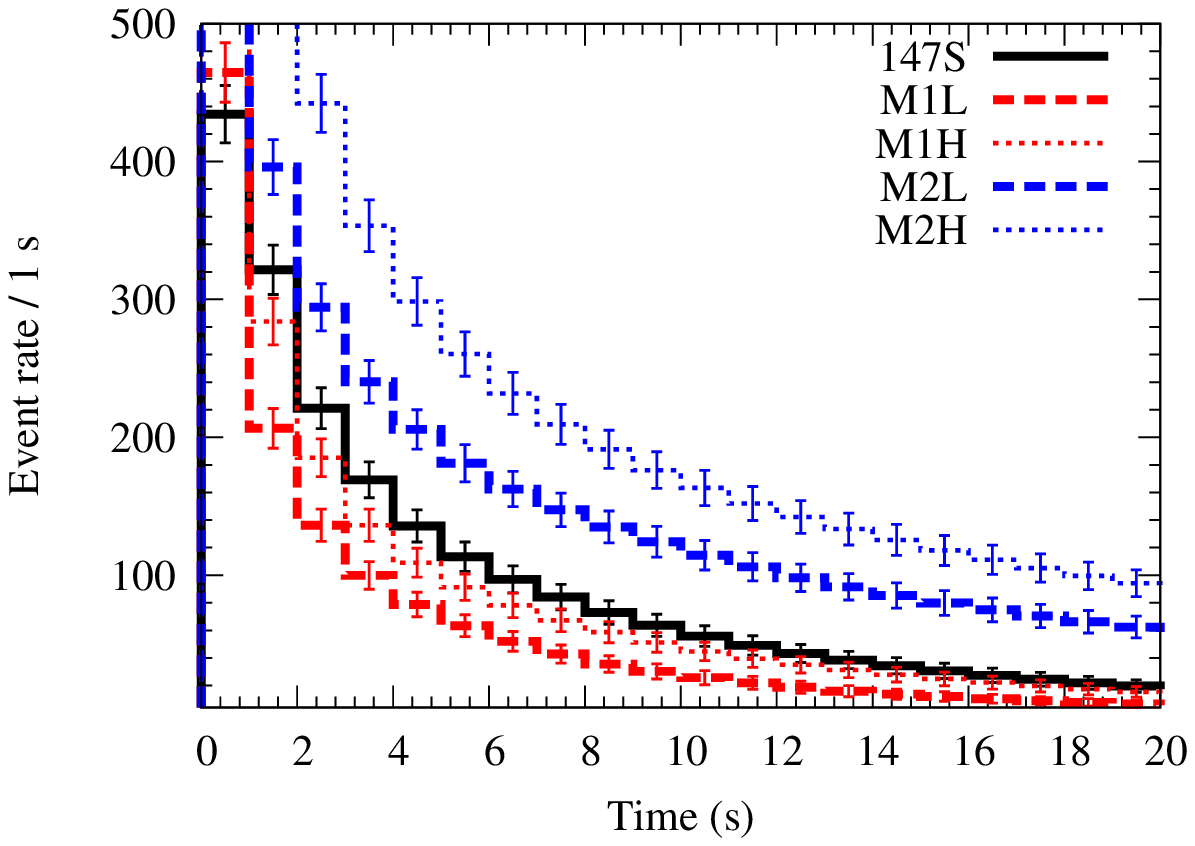}
\includegraphics[width=.45\textwidth]{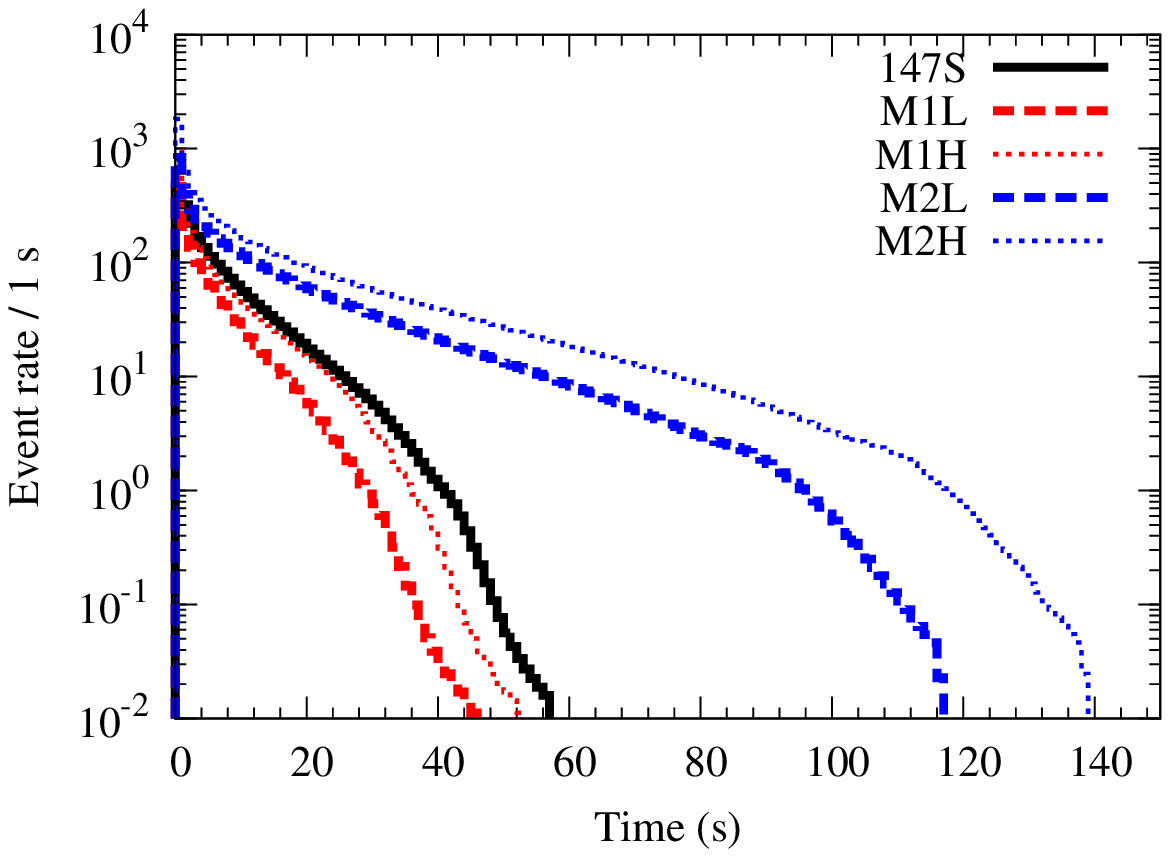}
\caption{Expected number of IBD events as a function of time from the birth of the PNS for a series of models described  in \S \ref{sec:PNSCdata}. Note that the time origin is different from the database model, since only the PNS cooling phase is calculated in these models. Error bars are not shown in the bottom panel for visibility. 
}
\label{fig:nulc-pns_event}
\end{figure}

We show in Figure \ref{fig:nulc-pns_event} the expected number of IBD events for the five PNS cooling models examined in \S \ref{sec:PNSCdata}.  
Event detection continues over 100 s for the case of PNSs with masses of 2.05M$_{\odot}$.  
The high-entropy model provides large event rates and persists up to 140 s as compared to the low-entropy model, due to their different thermal energies.  
The event rate drops faster in the case of the low mass (1.20M$_{\odot}$) model and in the fiducial model (1.35M$_{\odot}$), becoming undetectable at around 50 s.
The entropy profile of the low-entropy model is similar to the one in the fiducial model, therefore, the difference of the low-entropy and low-mass model from the fiducial model roughly reflects the difference of their PNS masses.  

\begin{figure}[tbp]
\includegraphics[width=0.45\textwidth]{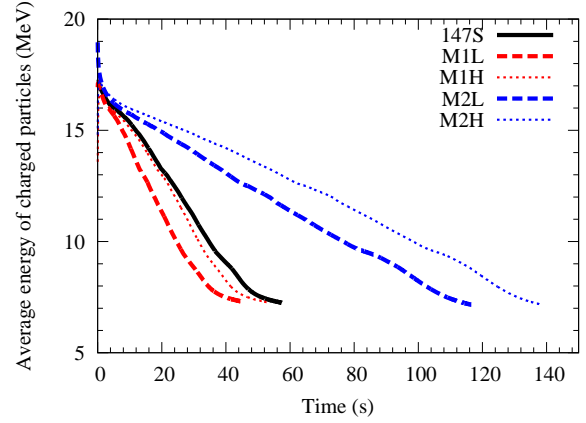}
\caption{Average energy of positrons from IBD reactions as a function of time for a Galactic supernova observed by Super-Kamiokande assuming an energy threshold of 5 MeV.}
\label{fig:average150s}
\end{figure}

Lastly, we discuss the evolution of event energies. 
Figure \ref{fig:average150s} shows the average energy of recoil positrons from IBD interactions of the neutrinos from models 147S, M1L, M1H, M2L, and M2H including the effect of the detector energy threshold. 
The positron energy is slightly higher than the average energy of the neutrinos shown in Figure \ref{fig:nulc-pns}. For instance, the final  energy of positrons for model 147S is 7.2 MeV, while that of the neutrinos is 3.7 MeV. This is because the positron's average energy is given by
\begin{align}
    \bar E_{e^+}=
    \frac
    {\int_{E_{\rm th}+\Delta}^{\infty}(\epsilon_\nu-\Delta)^5f_\nu(\epsilon_\nu-\Delta)d\epsilon_\nu}{\int_{E_{\rm th}+\Delta}^{\infty}(\epsilon_\nu-\Delta)^4f_\nu(\epsilon_\nu-\Delta)d\epsilon_\nu},
    \label{eq:bar_E_e}
\end{align}
where $E_{\rm th}$ is the threshold energy of positron detection at SK, $\epsilon_\nu$ is the energy of a neutrino that produces a positron, $\Delta=1.29$ MeV is the mass energy difference between neutrons and protons, and $f_\nu(\epsilon_\nu)$ is the neutrino phase space occupation function. 
Here, we assume the cross section of IBD is $\sigma(\epsilon_\nu)\propto \epsilon_\nu^2$, which gives the power of $\epsilon_\nu-\Delta$, and assume that the neutrino phase space occupation function is a Fermi-Dirac function without chemical potential, $f_\nu(\epsilon_\nu)=1/\left(1+e^{\epsilon_\nu/k_BT_\nu}\right)$ with the temperature $T_\nu$. 
In Tab. \ref{tab:positron_energy} we show the average energy of positrons with different neutrino temperatures and threshold energies. 
Thus, with a neutrino temperature $k_BT_\nu=\bar E_\nu/3.15=1.2$ MeV and $E_{\rm th}=$5 MeV, our average neutrino energy from the PNS cooling simulation and positron average energy are consistent.
By simply assuming $E_{\rm th}=\Delta=0$, we get $\bar E_{e^+}=5.07k_BT_\nu=1.61\bar E_\nu$, which is applicable for the early phase of neutrino emission from a hot PNS.

\begin{table}[]
    \centering
    \caption{Average energy of positrons from IBD reactions from Eq. \eqref{eq:bar_E_e}.}
    \begin{tabular}{l|ccc}
         & \multicolumn{3}{c}{$E_{\rm th}$} \\
         & 3 MeV & 5 MeV & 7 MeV \\
         \hline
         $kT_\nu=1$ MeV & 5.63 & 7.00 & 8.69 \\
         $kT_\nu=1.5$ MeV & 7.85 & 8.73 & 10.1 \\
         $kT_\nu=2$ MeV & 10.2 & 10.8 & 11.8 \\
         $kT_\nu=2.5$ MeV & 12.7 & 13.1 & 13.8 \\
         $kT_\nu=3$ MeV & 15.2 & 15.5 & 16.0 \\
         $kT_\nu=3.5$ MeV & 17.8 & 17.9 & 18.3 \\
         $kT_\nu=4$ MeV & 20.3 & 20.4 & 20.7 \\
         \hline
    \end{tabular}
    \label{tab:positron_energy}
\end{table}

\section{Time evolution and detection threshold}\label{sec:howlong}

\subsection{How long are the neutrinos detectable?}\label{sec:fadeout}

\begin{figure}[htbp]
\includegraphics[width=0.45\textwidth]{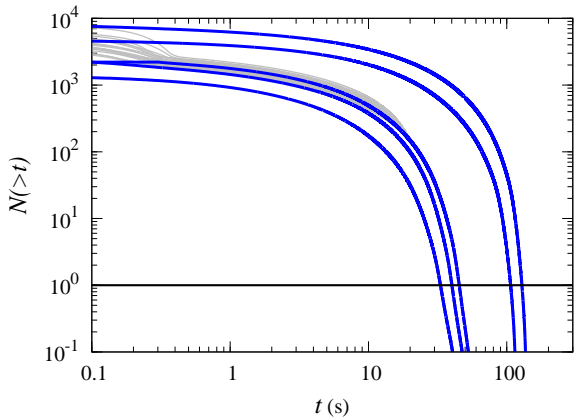}
\caption{Reverse cumulative event numbers as a function of time from PNS cooling calculations. Blue lines are the models from \S \ref{sec:PNSCdata} and gray lines are from \cite{naka13}.
}
\label{fig:cumulative_all}
\end{figure}

Here, we investigate the observable timescale of neutrinos from Galactic supernovae at 10 kpc. 
Figure~\ref{fig:cumulative_all} shows the reverse cumulative event number distributions of the models in \S \ref{sec:PNSCdata} (blue lines) and \cite{naka13} (gray lines).
Here the reverse cumulative event number is given by,
\begin{equation}
N(>t)=\int_t^\infty \dot N dt,
\end{equation}
where $\dot N$ is the event rate per unit time.
Since data from \cite{naka13} are only available up to 20 s, we use the event rate from model 147S at $t=20$ s in the gray lines. 
In the following we take the time when this cumulative event number is unity as the last observable time since the typical background rate of detectors after background reduction cuts is small compared to the expected supernova neutrino rate (see discussion in \S\ref{sec:summary}) implying that the current discussion is not influenced by systematic errors on the background.
For the canonical model (147S), the observable time is 45.3 s.  As it depends on the PNS mass, even for the smallest PNS mass observed so far \citep[$\approx 1.17$M$_\odot$, see][]{marti15} neutrinos can be observed for more than 30 s. More precisely, the observation time ranges from 33.2--40.1 s depending on the initial entropy.
For the most massive PNS presently known \cite[$\approx 2.0$M$_\odot$, see][]{antoni13} the range is 107--129 s.

\begin{figure}[htbp]
\includegraphics[width=0.45\textwidth]{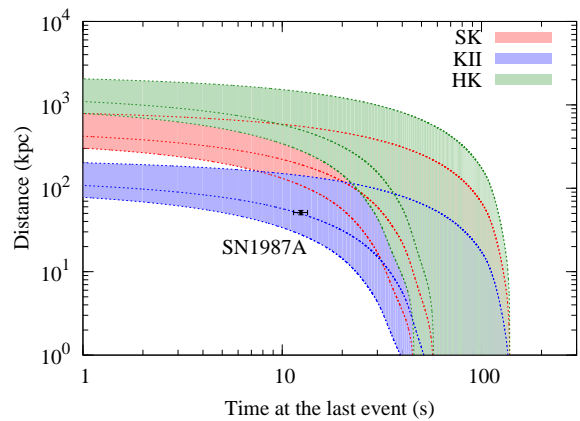}
\caption{Relationship between the observable time and distance to the supernova. Red, blue, and green shaded regions show Super-Kamiokande, Kamiokande-II, and Hyper-Kamiokande. The bottom, top and central lines in each band correspond to PNS models with low mass and small entropy, high mass and high entropy, and the canonical mass and entropy, respectively. SN 1987A is shown as a black point with errors of 1 s and 3.1 kpc and fits well within the Kamiokande-II region.}
\label{fig:distance-time}
\end{figure}

Figure \ref{fig:distance-time} gives the relationship between the observable timescale of neutrinos and the distance to the supernova. It is apparent that we can observe neutrinos for longer times for nearby supernovae. 
Colors show the dependence on the detector size; red shows the full volume of SK's inner detector (32.5 kton), blue shows that for Kamiokande-II (2.14~kton), and green is for Hyper-Kamiokande (220 kton). 
 Bands show the range for each detector assuming different models, where the lowest rate model has $M_{\rm NS,g}=1.20M_\odot$ and low initial entropy (M1L), while the highest has 2.05$M_\odot$ and high initial entropy (M2H) (see \S\ref{sec:PNSCdata}).
The black point gives values for SN 1987A, whose distance is 51.2$\pm$3.1 kpc \citep{pana91} and the observed duration was $\sim 12.4$ s \citep{hir87}.
This is consistent with the canonical model for Kamiokande-II, shown by the central dotted line in the blue region. 
Note that the total event number is also consistent with the observation at 11 or 12 events.

Note also that the current estimation is given assuming a kinetic energy threshold of 5 MeV, but the Kamiokande-II observation in \cite{hir87} used 7 MeV. 
Repeating the same calculation with their threshold, we find no significant difference from that with 5 MeV.

In this study, we employ the full 32.5 kton volume of the SK inner detector. Since the background level for this volume is highly uncertain, we perform the same calculations above with their standard fiducial volume, 22.5 kton, in order to investigate the impact of the detector size. The total event number is reduced to 69\% of the full volume and the observable time changes from 33.2 (M1L)--129 (M2H) s to 32.1--127 s. The change is small because near the time of the last event the event rates are rapidly decreasing (see Figs. \ref{fig:cumulative_all} and \ref{fig:distance-time}) and the reduced detector volume does not have a large impact on the last event, even though the total number of events is reduced. The background level in the fiducial volume is expected to be considerably lower than in the full volume and is negligibly small for times near the last event. Therefore, at least for an SN occurring within 10 kpc, SK will observe the last event without significant contamination from backgrounds.

\subsection{Backward time analysis}\label{sec:backward}

We propose a backward time analysis to explore the difference in models. 
It should be noted that the late time properties of the neutrino spectrum depend on a small number of parameters, which are completely different from those of the early epoch.
Whereas the late time evolution depends on the PNS mass, radius, and temperature, physics processes such as convection, SASI, mass accretion onto the PNS, and the onset of the explosion are necessary for modeling the neutrino light curve at early times.

\begin{figure}[tbp]
\includegraphics[width=0.45\textwidth]{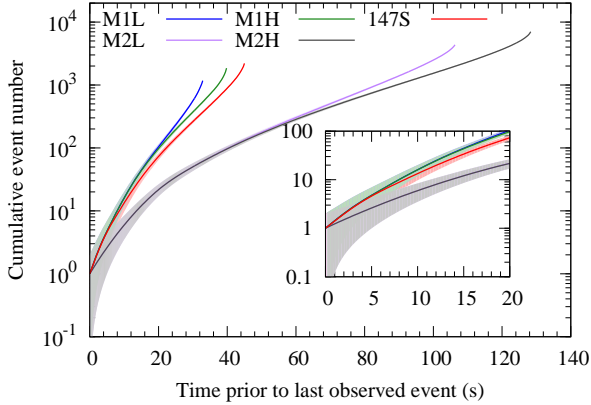}
\caption{Cumulative event number as a function of time measured backward from the expected last event. 
The shaded region shows the variation in the prediction assuming Poisson statistical uncertainties.
}
\label{fig:backward}
\end{figure}

Figure \ref{fig:backward} presents the cumulative event number as a function of time as measured backward in time from the last observed event (the time when $N(>t)=1$ as described above).
The shaded region shows the Poisson statistical uncertainty. 
It is clear that model groups with different PNS masses are well  separated in this metric (the M1L and M1H models have $M_{\rm NS,g}=1.20M_\odot$, while the M2L and M2H models have $M_{\rm NS,g}=2.06M_\odot$). 
This indicates that we can, in principle, infer the mass of the PNS formed by a supernova using the neutrino event count alone. Of course, the nuclear EOS is also an important ingredient characterizing the neutrino light curves and its impact will be discussed elsewhere (see \citealt{naka19}, for instance).

\begin{figure}[tbp]
\includegraphics[width=0.45\textwidth]{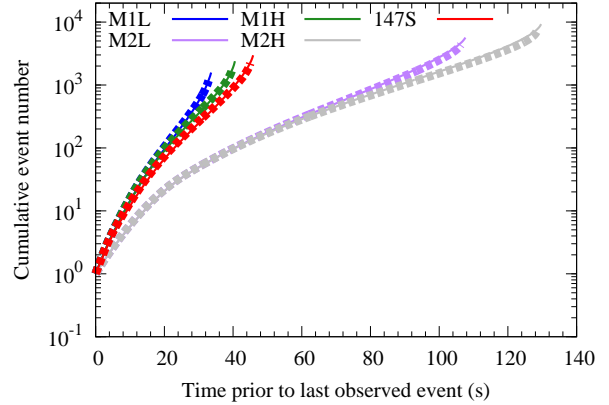}
\caption{Same as Figure \ref{fig:backward}, but comparing calculations with and without neutrino oscillation. 
Dashed lines are calculated using $\bar\nu_e$ directly and solid lines are calculated assuming $\nu_X$ is completely converted to $\bar\nu_e$. Note that the two lines are almost identical for the last $\mathcal{O}(100)$ events.
}
\label{fig:backward_osc}
\end{figure}

To investigate the impact of neutrino oscillations, we performed the same calculations exchanging $\bar\nu_e$ and $\nu_X$ completely. Though it is certainly an extreme scenario, reality should fall within the original calculation and this case. 
Figure \ref{fig:backward_osc} is the same as Figure \ref{fig:backward} but compares the calculations with (dashed lines) and without (solid lines) neutrino oscillations. The luminosity and spectra of $\bar\nu_e$ and $\nu_X$ are  similar at late times, so that the reverse cumulative event numbers for $t_{\rm back}\lesssim 20$~s  are roughly independent of these oscillations. 

\section{Data analysis strategy}
\label{sec:analysis}

In this section we summarize our strategy for analyzing data from the next Galactic SN neutrino burst.  
After the detection of a supernova burst, detailed detection data including the time, electron or positron energy, and possible direction of each observed neutrino event can be expected from Super-Kamiokande as was done in Table 1 of \cite{hir87} after the observation of SN 1987A. Based on this information we will analyze the data to extract the astrophysical details of the remnant in the following manner.

First, we make a gross evaluation from the total number of events during the burst.  
From Figure \ref{fig:sn_total_event}, the total number of IBD events provides a rough estimate of the distance to the SN up to a factor of three for the current range of our models.
The average energy of the charged particles can be used to extract the average energy of neutrinos and subsequently the temperature of neutrino-emitting object.  
Furthermore, the total energy radiated in neutrinos (assuming $\bar\nu_e$) can be used to infer the binding energy of the compact object.  
These values can be obtained in the same way as has been done for SN 1987A.  
The total energy depends on the distance, which can be obtained by  information from optical or measurements in other wavelengths.\footnote{SN 1987A was located in the LMC, enabling a high-precision distance measurement.} 
In the following analysis, we assume the distance is known to be 10 kpc.  

Next, we would like to utilize the time profile of the burst.  
Using the time sequence of the charged particle energies as time bins, we can extract the evolution of the neutrino event rate and average energy. The latter then provides the evolution of the neutrino temperature.  
Comparing the time profiles of these quantities with a set profiles calculated using models from the database can thus be used to infer the properties of the PNS as discussed in \S \ref{sec:event_number}.  However, it may be practically difficult to narrow down to an individual model due to the lack of information on the bounce time and complications arising from the impact of multi-dimensional hydrodynamic effects on the neutrino signal, which are closely related to the explosion mechanism.  

Therefore we take the backward time analysis discussed in Section \ref{sec:backward} in order to extract the astrophysical information and avoid this uncertainty.  
Producing a cumulative event rate backward in time from the last observed event and comparing with model expectations as in Figure \ref{fig:backward} allows for the extraction of information on the compact object formed in the supernova explosion.  
Furthermore, because the late phase of the neutrino burst is driven solely by emission from the cooling PNS and because the neutrino spectra among the different species are similar, we are free from oscillation effects.
Utilizing data from the late phase is simpler than the early phase and will help in the construction of a baseline for extracting information on the hydrodynamic instabilities such as SASI and convection.  

In order to complete the analysis with this strategy, we plan to provide a database using a fine grid of PNS gravitational masses and further investigate the dependence of the backward-in-time cumulative event distribution on the EOS.  
If we can successfully extract parameters of the compact object from the late phase neutrino signal, we may be able to further infer details of the explosion mechanism and characteristics of the progenitor using neutrinos in the early phase. 
Such additional studies will be reported elsewhere.

\section{Summary and discussion} \label{sec:summary}

Supernova neutrinos are essential to probe the final phase of massive star evolution. 
In particular, properties of the neutron star formed just after an explosion can be extracted from neutrino observations.  
To perform such an analysis for a Galactic event, we need a  comprehensive methodology to covering the entire timescale of neutrino emission. 
Although there are a number of modern simulations of neutrino-radiation hydrodynamics that focus on the early phase (less than 1 s after the bounce) emission, the late phase (more than 1 s after the bounce) has not yet been systematically studied.

In this study,  we investigated neutrino properties observable by Super-Kamiokande up until 20 s after the bounce using the database of \cite{naka13}. 
We also added five additional models by performing new PNS cooling calculations and studied the duration of observable neutrinos. We found that we will be able to observe neutrinos for more than 30 s even for a low-mass neutron star (gravitational mass of $1.20M_\odot$ ) and for more than 100 s for a high-mass neutron star ($2.05M_\odot$), assuming a supernova at 10 kpc.

In addition, we showed that the neutron star mass can be measured with the cumulative neutrino event distribution calculated as a function of time measured backwards from the last event. 
The neutrino oscillation effect was also investigated and found to have no influence on this metric because at late times the neutrino luminosities and spectra are almost flavor independent.

There are a few caveats. 
In this study, we employed only one nuclear EOS. 
As is well known, the details of the EOS are still under debate and can change the relationship between the neutron star mass and radius, changing the average and total neutrino energies.
We leave the EOS dependence for a future study, in which methods of resolving the degeneracy between the mass and radius from the neutrino signal will be discussed. 
Systematic errors from detailed neutrino interaction modeling and from the neutrino radiation transfer method during the final phase of PNS cooling will also be addressed in the next study.

Lastly, the background level of Super-Kamiokande needs to be further discussed.
In this paper, we neglect background contamination because of the small expected background rate in the fiducial volume for energies over 5 MeV. More specifically, for electrons and positrons, the measured background rate above 5 MeV of kinetic energy at SK was 150 per day at the time of GW150914 \citep{abe16}, which corresponds to 0.17 events during 100 s  in the 22.5 kton detector volume.
This property helps to determine the time of the last event. 
It should also be noted that this background estimate already takes into account reduction cuts \citep[for details, see][]{abe16b}. At energies above 5 MeV, the dominant background is expected to be from spallation products \citep{abe16}, whose rate can be estimated from measurements at SK \citep{Super-Kamiokande:2015xra} as roughly 2.5 events in 100 s, without the reduction cuts. 
Roughly speaking, the reduction cuts reduce the background level significantly (about a factor of 10), but also reduce the neutrino signal by $\sim$20\% \citep{abe16b}. 
Though background estimates in the full SK volume are currently unavailable, in the event of a real supernova we anticipate the actual analysis will use as much of the volume as the neutrino event rate allows and will transition to a smaller volume with tighter analysis cuts and lower backgrounds to extract the last few supernova neutrino events.  
Therefore, in this work we imagine a two stage analysis.  
First, since the total number of events will be dominated by the supernova signal, it can be obtained without applying reduction cuts.
Second, the timing of the last event should be obtained using the reduction cuts to reduce contamination from spallation products.
Once the last event's time is known, the backward analysis described above can be performed on the data set without the reduction cuts because the expected contamination is small (cf. two background events compared to $\mathcal{O}(10^3)$ in 100 s). 
Note that the spallation background rate should scale with volume, such that above 5 MeV this argument should hold even in the analysis of the full SK volume.

For the larger volume or for Hyper-Kamiokande, significantly larger background levels may be harmful to a precise determination. 
For those cases a fit to the data using a model of the neutrino light curve is need. In \cite{halz09}, the authors proposed a method to reconstruct the {\it onset} of neutrino emission and their model could be useful for this purpose. Naively, until the time when the noise level (Poisson fluctuations of the average background rate) is sufficiently lower than the real event rate by, roughly speaking, a factor of three, such fitting is possible (for real event rates, see the bottom panel of Figure \ref{fig:nulc-pns_event}). More detailed studies will be reported in a forthcoming paper.

\acknowledgments

This work is supported by
Grant-in-Aid for Scientific Research
(15K05093, 16K17665,  17H02864)
and 
Grant-in-Aid for Scientific Research on Innovative areas 
(26104006, 17H05203, 17H06357, 17H06365, 18H04586, 18H05437)
from the Ministry of Education, Culture, Sports, Science and Technology (MEXT), Japan.

For providing high performance computing resources, 
Research Center for Nuclear Physics in Osaka University, 
Yukawa Institute of Theoretical Physics in Kyoto University, 
Center for Computational Astrophysics in National Astronomical Observatory of Japan,
Computing Research Center in KEK, 
JLDG on SINET4 of NII, 
Information Technology Center in Nagoya University, 
and 
Information Technology Center in University of Tokyo 
are acknowledged.  

This work was partly supported by 
research programs at K-computer of the RIKEN AICS, 
HPCI Strategic Program of Japanese MEXT, 
“Priority Issue on Post-K computer” (Elucidation of the Fundamental Laws and Evolution of the Universe)
and 
Joint Institute for Computational Fundamental Sciences (JICFus).


\begin{thebibliography}{}
\expandafter\ifx\csname natexlab\endcsname\relax\def\natexlab#1{#1}\fi
\providecommand{\url}[1]{\href{#1}{#1}}

\bibitem[{{Abe} {et~al.}(2011){Abe}, {Abe}, {Aihara}, {Fukuda}, {Hayato},
  {Huang}, {Ichikawa}, {Ikeda}, {Inoue}, {Ishino}, {Itow}, {Kajita}, {Kameda},
  {Kishimoto}, {Koga}, {Koshio}, {Lee}, {Minamino}, {Miura}, {Moriyama},
  {Nakahata}, {Nakamura}, {Nakaya}, {Nakayama}, {Nishijima}, {Nishimura},
  {Obayashi}, {Okumura}, {Sakuda}, {Sekiya}, {Shiozawa}, {Suzuki}, {Suzuki},
  {Takeda}, {Takeuchi}, {Tanaka}, {Tasaka}, {Tomura}, {Vagins}, {Wang}, \&
  {Yokoyama}}]{hypk11}
{Abe}, K., {Abe}, T., {Aihara}, H., {et~al.} 2011, ArXiv e-prints,
  arXiv:1109.3262

\bibitem[{{Abe} {et~al.}(2016{\natexlab{a}}){Abe}, {Haga}, {Hayato}, {Ikeda},
  {Iyogi}, {Kameda}, {Kishimoto}, {Miura}, {Moriyama}, {Nakahata}, {Nakajima},
  {Nakano}, {Nakayama}, {Orii}, {Sekiya}, {Shiozawa}, {Takeda}, {Tanaka},
  {Tasaka}, {Tomura}, {Akutsu}, {Kajita}, {Kaneyuki}, {Nishimura}, {Richard},
  {Okumura}, {Labarga}, {Fernandez}, {Blaszczyk}, {Gustafson}, {Kachulis},
  {Kearns}, {Raaf}, {Stone}, {Sulak}, {Berkman}, {Nantais}, {Tobayama},
  {Goldhaber}, {Kropp}, {Mine}, {Weatherly}, {Smy}, {Sobel}, {Takhistov},
  {Ganezer}, {Hartfiel}, {Hill}, {Hong}, {Kim}, {Lim}, {Park}, {Himmel}, {Li},
  {O'Sullivan}, {Scholberg}, {Walter}, {Ishizuka}, {Nakamura}, {Jang}, {Choi},
  {Learned}, {Matsuno}, {Smith}, {Friend}, {Hasegawa}, {Ishida}, {Ishii},
  {Kobayashi}, {Nakadaira}, {Nakamura}, {Oyama}, {Sakashita}, {Sekiguchi},
  {Tsukamoto}, {Suzuki}, {Takeuchi}, {Yano}, {Cao}, {Hiraki}, {Hirota},
  {Huang}, {Jiang}, {Minamino}, {Nakaya}, {Patel}, {Wendell}, {Suzuki},
  {Fukuda}, {Itow}, {Suzuki}, {Mijakowski}, {Frankiewicz}, {Hignight}, {Imber},
  {Jung}, {Li}, {Palomino}, {Santucci}, {Wilking}, {Yanagisawa}, {Fukuda},
  {Ishino}, {Kayano}, {Kibayashi}, {Koshio}, {Mori}, {Sakuda}, {Xu}, {Kuno},
  {Tacik}, {Kim}, {Okazawa}, {Choi}, {Nishijima}, {Koshiba}, {Totsuka}, {Suda},
  {Yokoyama}, {Bronner}, {Calland}, {Hartz}, {Martens}, {Marti}, {Suzuki},
  {Vagins}, {Martin}, {Tanaka}, {Konaka}, {Chen}, {Wan}, {Zhang}, {Wilkes}, \&
  {Super-Kamiokande Collaboration}}]{abe16}
{Abe}, K., {Haga}, K., {Hayato}, Y., {et~al.} 2016{\natexlab{a}}, \apj, 830,
  L11

\bibitem[{{Abe} {et~al.}(2016{\natexlab{b}}){Abe}, {Haga}, {Hayato}, {Ikeda},
  {Iyogi}, {Kameda}, {Kishimoto}, {Marti}, {Miura}, {Moriyama}, {Nakahata},
  {Nakajima}, {Nakayama}, {Orii}, {Sekiya}, {Shiozawa}, {Sonoda}, {Takeda},
  {Tanaka}, {Takenaga}, {Tasaka}, {Tomura}, {Ueno}, {Yokozawa}, {Akutsu},
  {Irvine}, {Kaji}, {Kajita}, {Kametani}, {Kaneyuki}, {Lee}, {Nishimura},
  {McLachlan}, {Okumura}, {Richard}, {Labarga}, {Fernandez}, {Blaszczyk},
  {Gustafson}, {Kachulis}, {Kearns}, {Raaf}, {Stone}, {Sulak}, {Berkman},
  {Tobayama}, {Goldhaber}, {Bays}, {Carminati}, {Griskevich}, {Kropp}, {Mine},
  {Renshaw}, {Smy}, {Sobel}, {Takhistov}, {Weatherly}, {Ganezer}, {Hartfiel},
  {Hill}, {Keig}, {Hong}, {Kim}, {Lim}, {Park}, {Akiri}, {Albert}, {Himmel},
  {Li}, {O'Sullivan}, {Scholberg}, {Walter}, {Wongjirad}, {Ishizuka},
  {Nakamura}, {Jang}, {Choi}, {Learned}, {Matsuno}, {Smith}, {Friend},
  {Hasegawa}, {Ishida}, {Ishii}, {Kobayashi}, {Nakadaira}, {Nakamura},
  {Nishikawa}, {Oyama}, {Sakashita}, {Sekiguchi}, {Tsukamoto}, {Nakano},
  {Suzuki}, {Takeuchi}, {Yano}, {Cao}, {Hayashino}, {Hiraki}, {Hirota},
  {Huang}, {Ieki}, {Jiang}, {Kikawa}, {Minamino}, {Murakami}, {Nakaya},
  {Patel}, {Suzuki}, {Takahashi}, {Wendell}, {Fukuda}, {Itow}, {Mitsuka},
  {Muto}, {Suzuki}, {Mijakowski}, {Frankiewicz}, {Hignight}, {Imber}, {Jung},
  {Li}, {Palomino}, {Santucci}, {Taylor}, {Vilela}, {Wilking}, {Yanagisawa},
  {Fukuda}, {Ishino}, {Kayano}, {Kibayashi}, {Koshio}, {Mori}, {Sakuda},
  {Takeuchi}, {Yamaguchi}, {Kuno}, {Tacik}, {Kim}, {Okazawa}, {Choi}, {Ito},
  {Nishijima}, {Koshiba}, {Totsuka}, {Suda}, {Yokoyama}, {Bronner}, {Calland},
  {Hartz}, {Martens}, {Obayashi}, {Suzuki}, {Vagins}, {Nantais}, {Martin}, {de
  Perio}, {Tanaka}, {Konaka}, {Chen}, {Sui}, {Wan}, {Yang}, {Zhang}, {Zhang},
  {Connolly}, {Dziomba}, \& {Wilkes}}]{abe16b}
{Abe}, K., {Haga}, Y., {Hayato}, Y., {et~al.} 2016{\natexlab{b}}, \prd, 94,
  052010

\bibitem[{{Abe} {et~al.}(2018){Abe}, {Abe}, {Aihara}, {Aimi}, {Akutsu},
  {Andreopoulos}, {Anghel}, {Anthony}, \& et~al.}]{hypk16}
{Abe}, K., {Abe}, K., {Aihara}, H., {et~al.} 2018, ArXiv e-prints,
  arXiv:1805.04163

\bibitem[{{Antoniadis} {et~al.}(2013){Antoniadis}, {Freire}, {Wex}, {Tauris},
  {Lynch}, {van Kerkwijk}, {Kramer}, {Bassa}, {Dhillon}, {Driebe}, {Hessels},
  {Kaspi}, {Kondratiev}, {Langer}, {Marsh}, {McLaughlin}, {Pennucci}, {Ransom},
  {Stairs}, {van Leeuwen}, {Verbiest}, \& {Whelan}}]{antoni13}
{Antoniadis}, J., {Freire}, P.~C.~C., {Wex}, N., {et~al.} 2013, Science, 340,
  448

\bibitem[{{Baumgarte} {et~al.}(1996){Baumgarte}, {Janka}, {Keil}, {Shapiro}, \&
  {Teukolsky}}]{bau96b}
{Baumgarte}, T.~W., {Janka}, H.-T., {Keil}, W., {Shapiro}, S.~L., \&
  {Teukolsky}, S.~A. 1996, \apj, 468, 823

\bibitem[{{Beacom} \& {Vagins}(2004)}]{bea04}
{Beacom}, J.~F., \& {Vagins}, M.~R. 2004, Physical Review Letters, 93, 171101

\bibitem[{Bionta {et~al.}(1987)}]{bio87}
Bionta, R.~M., {et~al.} 1987, Phys.\ Rev.\ Lett., 58, 1494

\bibitem[{{Bludman} \& {Schinder}(1988)}]{blu88}
{Bludman}, S.~A., \& {Schinder}, P.~J. 1988, \apj, 326, 265

\bibitem[{{Brdar} {et~al.}(2018){Brdar}, {Lindner}, \& {Xu}}]{brda18}
{Brdar}, V., {Lindner}, M., \& {Xu}, X.-J. 2018, Journal of Cosmology and
  Astro-Particle Physics, 2018, 025

\bibitem[{{Burrows}(1988)}]{bur88}
{Burrows}, A. 1988, \apj, 334, 891

\bibitem[{{Burrows}(2013)}]{bur13}
---. 2013, Reviews of Modern Physics, 85, 245

\bibitem[{{Burrows} \& {Lattimer}(1986)}]{bur86}
{Burrows}, A., \& {Lattimer}, J.~M. 1986, \apj, 307, 178

\bibitem[{{Burrows} \& {Lattimer}(1987)}]{bur87}
---. 1987, \apjl, 318, L63

\bibitem[{{Camelio} {et~al.}(2017){Camelio}, {Lovato}, {Gualtieri}, {Benhar},
  {Pons}, \& {Ferrari}}]{cam17}
{Camelio}, G., {Lovato}, A., {Gualtieri}, L., {et~al.} 2017, \prd, 96, 043015

\bibitem[{{Dalhed} {et~al.}(1999){Dalhed}, {Wilson}, \& {Mayle}}]{dal99}
{Dalhed}, H.~E., {Wilson}, J.~R., \& {Mayle}, R.~W. 1999, Nuclear Physics B
  Proceedings Supplements, 77, 429

\bibitem[{{Demorest} {et~al.}(2010){Demorest}, {Pennucci}, {Ransom}, {Roberts},
  \& {Hessels}}]{demo10}
{Demorest}, P.~B., {Pennucci}, T., {Ransom}, S.~M., {Roberts}, M.~S.~E., \&
  {Hessels}, J.~W.~T. 2010, \nat, 467, 1081

\bibitem[{{Fischer}(2016)}]{fis16b}
{Fischer}, T. 2016, \aap, 593, A103

\bibitem[{{Fischer} {et~al.}(2010){Fischer}, {Whitehouse}, {Mezzacappa},
  {Thielemann}, \& {Liebend{\"o}rfer}}]{fis10}
{Fischer}, T., {Whitehouse}, S.~C., {Mezzacappa}, A., {Thielemann}, F.-K., \&
  {Liebend{\"o}rfer}, M. 2010, \aap, 517, A80

\bibitem[{{Fukuda} {et~al.}(2003){Fukuda}, {Fukuda}, {Hayakawa}, {Ichihara},
  {Itow}, {Kajita}, {Kameda}, {Kaneyuki}, {Kasuga}, {Kobayashi}, {Kobayashi},
  {Koshio}, {Miura}, {Moriyama}, {Nakahata}, {Nakayama}, {Namba}, {Obayashi},
  {Okada}, {Oketa}, {Okumura}, {Oyabu}, {Sakurai}, {Shiozawa}, {Suzuki},
  {Takeuchi}, {Toshito}, {Totsuka}, {Yamada}, {Desai}, {Earl}, {Hong},
  {Kearns}, {Masuzawa}, {Messier}, {Stone}, {Sulak}, {Walter}, {Wang},
  {Scholberg}, {Barszczak}, {Casper}, {Liu}, {Gajewski}, {Halverson}, {Hsu},
  {Kropp}, {Mine}, {Price}, {Reines}, {Smy}, {Sobel}, {Vagins}, {Ganezer},
  {Keig}, {Ellsworth}, {Tasaka}, {Flanagan}, {Kibayashi}, {Learned}, {Matsuno},
  {Stenger}, {Hayato}, {Ishii}, {Ichikawa}, {Kanzaki}, {Kobayashi}, {Maruyama},
  {Nakamura}, {Oyama}, {Sakai}, {Sakuda}, {Sasaki}, {Echigo}, {Iwashita},
  {Kohama}, {Suzuki}, {Hasegawa}, {Inagaki}, {Kato}, {Maesaka}, {Nakaya},
  {Nishikawa}, {Yamamoto}, {Haines}, {Kim}, {Sanford}, {Svoboda}, {Blaufuss},
  {Chen}, {Conner}, {Goodman}, {Guillian}, {Sullivan}, {Turcan}, {Habig},
  {Ackerman}, {Goebel}, {Hill}, {Jung}, {Kato}, {Kerr}, {Malek}, {Martens},
  {Mauger}, {McGrew}, {Sharkey}, {Viren}, {Yanagisawa}, {Doki}, {Inaba}, {Ito},
  {Kirisawa}, {Kitaguchi}, {Mitsuda}, {Miyano}, {Saji}, {Takahata},
  {Takahashi}, {Higuchi}, {Kajiyama}, {Kusano}, {Nagashima}, {Nitta}, {Takita},
  {Yamaguchi}, {Yoshida}, {Kim}, {Kim}, {Yoo}, {Okazawa}, {Etoh}, {Fujita},
  {Gando}, {Hasegawa}, {Hasegawa}, {Hatakeyama}, {Inoue}, {Ishihara},
  {Iwamoto}, {Koga}, {Nishiyama}, {Ogawa}, {Shirai}, {Suzuki}, {Takayama},
  {Tsushima}, {Koshiba}, {Ichikawa}, {Hashimoto}, {Hatakeyama}, {Koike},
  {Horiuchi}, {Nemoto}, {Nishijima}, {Takeda}, {Fujiyasu}, {Futagami},
  {Ishino}, {Kanaya}, {Morii}, {Nishihama}, {Nishimura}, {Suzuki}, {Watanabe},
  {Kielczewska}, {Golebiewska}, {Berns}, {Boyd}, {Doyle}, {George}, {Stachyra},
  {Wai}, {Wilkes}, {Young}, {Kobayashi}, \& {Super-Kamiokande
  Collaboration}}]{2003NIMPA.501..418F}
{Fukuda}, S., {Fukuda}, Y., {Hayakawa}, T., {et~al.} 2003, Nuclear Instruments
  and Methods in Physics Research A, 501, 418

\bibitem[{{Halzen} \& {Raffelt}(2009)}]{halz09}
{Halzen}, F., \& {Raffelt}, G.~G. 2009, \prd, 80, 087301

\bibitem[{Hirata {et~al.}(1987)}]{hir87}
Hirata, K., {et~al.} 1987, Phys.\ Rev.\ Lett., 58, 1490

\bibitem[{{Horiuchi} \& {Kneller}(2018)}]{hor18b}
{Horiuchi}, S., \& {Kneller}, J.~P. 2018, Journal of Physics G Nuclear Physics,
  45, 043002

\bibitem[{{Horiuchi} {et~al.}(2018){Horiuchi}, {Sumiyoshi}, {Nakamura},
  {Fischer}, {Summa}, {Takiwaki}, {Janka}, \& {Kotake}}]{hor18a}
{Horiuchi}, S., {Sumiyoshi}, K., {Nakamura}, K., {et~al.} 2018, \mnras, 475,
  1363

\bibitem[{{H{\"u}depohl} {et~al.}(2010){H{\"u}depohl}, {M{\"u}ller}, {Janka},
  {Marek}, \& {Raffelt}}]{hue10}
{H{\"u}depohl}, L., {M{\"u}ller}, B., {Janka}, H.-T., {Marek}, A., \&
  {Raffelt}, G.~G. 2010, Physical Review Letters, 104, 251101

\bibitem[{{Ikeda} {et~al.}(2007){Ikeda}, {Takeda}, {Fukuda}, {Vagins}, {Abe},
  {Iida}, {Ishihara}, {Kameda}, {Koshio}, {Minamino}, {Mitsuda}, {Miura},
  {Moriyama}, {Nakahata}, {Obayashi}, {Ogawa}, {Sekiya}, {Shiozawa}, {Suzuki},
  {Takeuchi}, {Ueshima}, {Watanabe}, {Yamada}, {Higuchi}, {Ishihara},
  {Ishitsuka}, {Kajita}, {Kaneyuki}, {Mitsuka}, {Nakayama}, {Nishino},
  {Okumura}, {Saji}, {Takenaga}, {Clark}, {Desai}, {Dufour}, {Kearns},
  {Likhoded}, {Litos}, {Raaf}, {Stone}, {Sulak}, {Wang}, {Goldhaber}, {Casper},
  {Cravens}, {Dunmore}, {Kropp}, {Liu}, {Mine}, {Regis}, {Smy}, {Sobel},
  {Ganezer}, {Hill}, {Keig}, {Jang}, {Kim}, {Lim}, {Scholberg}, {Tanimoto},
  {Walter}, {Wendell}, {Ellsworth}, {Tasaka}, {Guillian}, {Learned}, {Matsuno},
  {Messier}, {Hayato}, {Ichikawa}, {Ishida}, {Ishii}, {Iwashita}, {Kobayashi},
  {Nakadaira}, {Nakamura}, {Nitta}, {Oyama}, {Totsuka}, {Suzuki}, {Hasegawa},
  {Hiraide}, {Maesaka}, {Nakaya}, {Nishikawa}, {Sasaki}, {Yamamoto},
  {Yokoyama}, {Haines}, {Dazeley}, {Hatakeyama}, {Svoboda}, {Sullivan},
  {Turcan}, {Habig}, {Sato}, {Itow}, {Koike}, {Tanaka}, {Jung}, {Kato},
  {Kobayashi}, {Malek}, {McGrew}, {Sarrat}, {Terri}, {Yanagisawa}, {Tamura},
  {Idehara}, {Sakuda}, {Sugihara}, {Kuno}, {Yoshida}, {Kim}, {Yang}, {Yoo},
  {Ishizuka}, {Okazawa}, {Choi}, {Seo}, {Gando}, {Hasegawa}, {Inoue}, {Furuse},
  {Ishii}, {Nishijima}, {Ishino}, {Watanabe}, {Koshiba}, {Chen}, {Deng}, {Liu},
  {Kielczewska}, {Zalipska}, {Berns}, {Gran}, {Shiraishi}, {Stachyra},
  {Thrane}, {Washburn}, {Wilkes}, \& {Super-KAMIOKANDE Collaboration}}]{ike07}
{Ikeda}, M., {Takeda}, A., {Fukuda}, Y., {et~al.} 2007, \apj, 669, 519

\bibitem[{{Janka}(2017{\natexlab{a}})}]{jan17bk1}
{Janka}, H.-T. 2017{\natexlab{a}}, {Neutrino-Driven Explosions}, ed. A.~W.
  {Alsabti} \& P.~{Murdin}, 1095

\bibitem[{{Janka}(2017{\natexlab{b}})}]{jan17bk2}
---. 2017{\natexlab{b}}, {Neutrino Emission from Supernovae}, ed. A.~W.
  {Alsabti} \& P.~{Murdin}, 1575

\bibitem[{{Janka} \& {Hillebrandt}(1989)}]{jan89}
{Janka}, H.-T., \& {Hillebrandt}, W. 1989, \aap, 224, 49

\bibitem[{{Janka} {et~al.}(2016){Janka}, {Melson}, \& {Summa}}]{jan16}
{Janka}, H.-T., {Melson}, T., \& {Summa}, A. 2016, Annual Review of Nuclear and
  Particle Science, 66, 341

\bibitem[{{Kato} {et~al.}(2017){Kato}, {Nagakura}, {Furusawa}, {Takahashi},
  {Umeda}, {Yoshida}, {Ishidoshiro}, \& {Yamada}}]{kat17}
{Kato}, C., {Nagakura}, H., {Furusawa}, S., {et~al.} 2017, \apj, 848, 48

\bibitem[{{Keil} \& {Janka}(1995)}]{kei95}
{Keil}, W., \& {Janka}, H.-T. 1995, \aap, 296, 145

\bibitem[{{Koshiba}(1992)}]{kos92}
{Koshiba}, M. 1992, \physrep, 220, 229

\bibitem[{{Kotake} {et~al.}(2006){Kotake}, {Sato}, \& {Takahashi}}]{kot06}
{Kotake}, K., {Sato}, K., \& {Takahashi}, K. 2006, Reports on Progress in
  Physics, 69, 971

\bibitem[{{Kotake} {et~al.}(2012){Kotake}, {Sumiyoshi}, {Yamada}, {Takiwaki},
  {Kuroda}, {Suwa}, \& {Nagakura}}]{kot12}
{Kotake}, K., {Sumiyoshi}, K., {Yamada}, S., {et~al.} 2012, Progress of
  Theoretical and Experimental Physics, 2012, 010000

\bibitem[{{Kuroda} {et~al.}(2017){Kuroda}, {Kotake}, {Hayama}, \&
  {Takiwaki}}]{kur17}
{Kuroda}, T., {Kotake}, K., {Hayama}, K., \& {Takiwaki}, T. 2017, \apj, 851, 62

\bibitem[{{Lund} {et~al.}(2012){Lund}, {Wongwathanarat}, {Janka}, {M{\"u}ller},
  \& {Raffelt}}]{lun12}
{Lund}, T., {Wongwathanarat}, A., {Janka}, H.-T., {M{\"u}ller}, E., \&
  {Raffelt}, G. 2012, \prd, 86, 105031

\bibitem[{{Marek} {et~al.}(2009){Marek}, {Janka}, \& {M{\"u}ller}}]{mar09}
{Marek}, A., {Janka}, H.-T., \& {M{\"u}ller}, E. 2009, \aap, 496, 475

\bibitem[{{Martinez} {et~al.}(2015){Martinez}, {Stovall}, {Freire}, {Deneva},
  {Jenet}, {McLaughlin}, {Bagchi}, {Bates}, \& {Ridolfi}}]{marti15}
{Martinez}, J.~G., {Stovall}, K., {Freire}, P.~C.~C., {et~al.} 2015, \apj, 812,
  143

\bibitem[{{Mart{\'{\i}}nez-Pinedo} {et~al.}(2012){Mart{\'{\i}}nez-Pinedo},
  {Fischer}, {Lohs}, \& {Huther}}]{pin12}
{Mart{\'{\i}}nez-Pinedo}, G., {Fischer}, T., {Lohs}, A., \& {Huther}, L. 2012,
  Physical Review Letters, 109, 251104

\bibitem[{{Mayle} {et~al.}(1987){Mayle}, {Wilson}, \& {Schramm}}]{may87}
{Mayle}, R., {Wilson}, J.~R., \& {Schramm}, D.~N. 1987, \apj, 318, 288

\bibitem[{{Mirizzi} {et~al.}(2016){Mirizzi}, {Tamborra}, {Janka}, {Saviano},
  {Scholberg}, {Bollig}, {H{\"u}depohl}, \& {Chakraborty}}]{mir16}
{Mirizzi}, A., {Tamborra}, I., {Janka}, H.-T., {et~al.} 2016, Nuovo Cimento
  Rivista Serie, 39, 1

\bibitem[{{Myra} \& {Burrows}(1990)}]{myr90}
{Myra}, E.~S., \& {Burrows}, A. 1990, \apj, 364, 222

\bibitem[{{Nakamura} {et~al.}(2015){Nakamura}, {Takiwaki}, {Kuroda}, \&
  {Kotake}}]{nak15}
{Nakamura}, K., {Takiwaki}, T., {Kuroda}, T., \& {Kotake}, K. 2015, \pasj, 67,
  107

\bibitem[{{Nakazato} {et~al.}(2013){Nakazato}, {Sumiyoshi}, {Suzuki}, {Totani},
  {Umeda}, \& {Yamada}}]{naka13}
{Nakazato}, K., {Sumiyoshi}, K., {Suzuki}, H., {et~al.} 2013, \apjs, 205, 2

\bibitem[{{Nakazato} \& {Suzuki}(2019)}]{naka19}
{Nakazato}, K., \& {Suzuki}, H. 2019, The Astrophysical Journal, 878, 25

\bibitem[{{Nakazato} {et~al.}(2018){Nakazato}, {Suzuki}, \& {Togashi}}]{nak18}
{Nakazato}, K., {Suzuki}, H., \& {Togashi}, H. 2018, \prc, 97, 035804

\bibitem[{{O'Connor} \& {Ott}(2013)}]{oco13}
{O'Connor}, E., \& {Ott}, C.~D. 2013, \apj, 762, 126

\bibitem[{{Odrzywolek} {et~al.}(2004){Odrzywolek}, {Misiaszek}, \&
  {Kutschera}}]{odr04}
{Odrzywolek}, A., {Misiaszek}, M., \& {Kutschera}, M. 2004, Astroparticle
  Physics, 21, 303

\bibitem[{{Panagia} {et~al.}(1991){Panagia}, {Gilmozzi}, {Macchetto}, {Adorf},
  \& {Kirshner}}]{pana91}
{Panagia}, N., {Gilmozzi}, R., {Macchetto}, F., {Adorf}, H.-M., \& {Kirshner},
  R.~P. 1991, \apjl, 380, L23

\bibitem[{Pons {et~al.}(2001{\natexlab{a}})Pons, Miralles, Prakash, \&
  Lattimer}]{pon01b}
Pons, J.~A., Miralles, J.~A., Prakash, M., \& Lattimer, J.~M.
  2001{\natexlab{a}}, Astrophys.\ J., 553, 382

\bibitem[{Pons {et~al.}(1999)Pons, Reddy, Prakash, Lattimer, \&
  Miralles}]{pon99}
Pons, J.~A., Reddy, S., Prakash, M., Lattimer, J.~M., \& Miralles, J.~A. 1999,
  Astrophys.\ J., 513, 780

\bibitem[{Pons {et~al.}(2001{\natexlab{b}})Pons, Steiner, Prakash, \&
  Lattimer}]{pon01a}
Pons, J.~A., Steiner, A.~W., Prakash, M., \& Lattimer, J.~M.
  2001{\natexlab{b}}, Phys.\ Rev.\ Lett., 86, 5223

\bibitem[{{Roberts}(2012)}]{rob12b}
{Roberts}, L.~F. 2012, \apj, 755, 126

\bibitem[{{Roberts} {et~al.}(2012){Roberts}, {Shen}, {Cirigliano}, {Pons},
  {Reddy}, \& {Woosley}}]{rob12a}
{Roberts}, L.~F., {Shen}, G., {Cirigliano}, V., {et~al.} 2012, Physical Review
  Letters, 108, 061103

\bibitem[{{Sato} \& {Suzuki}(1987{\natexlab{a}})}]{sat87a}
{Sato}, K., \& {Suzuki}, H. 1987{\natexlab{a}}, Physical Review Letters, 58,
  2722

\bibitem[{{Sato} \& {Suzuki}(1987{\natexlab{b}})}]{sat87b}
---. 1987{\natexlab{b}}, Physics Letters B, 196, 267

\bibitem[{{Scholberg}(2012)}]{sch12}
{Scholberg}, K. 2012, Annual Review of Nuclear and Particle Science, 62, 81

\bibitem[{{Shen} {et~al.}(1998{\natexlab{a}}){Shen}, {Toki}, {Oyamatsu}, \&
  {Sumiyoshi}}]{she98a}
{Shen}, H., {Toki}, H., {Oyamatsu}, K., \& {Sumiyoshi}, K. 1998{\natexlab{a}},
  Nuclear Physics A, 637, 435

\bibitem[{{Shen} {et~al.}(1998{\natexlab{b}}){Shen}, {Toki}, {Oyamatsu}, \&
  {Sumiyoshi}}]{she98b}
---. 1998{\natexlab{b}}, Progress of Theoretical Physics, 100, 1013

\bibitem[{{Shen} {et~al.}(2011){Shen}, {Toki}, {Oyamatsu}, \&
  {Sumiyoshi}}]{she11}
---. 2011, Astrophys.\ J.\ Suppl., 197, 20

\bibitem[{{Strumia} \& {Vissani}(2003)}]{stru03}
{Strumia}, A., \& {Vissani}, F. 2003, Physics Letters B, 564, 42

\bibitem[{{Sumiyoshi} {et~al.}(1995){Sumiyoshi}, {Suzuki}, \& {Toki}}]{sum95}
{Sumiyoshi}, K., {Suzuki}, H., \& {Toki}, H. 1995, \aap, 303, 475

\bibitem[{Sumiyoshi {et~al.}(2007)Sumiyoshi, Yamada, \& Suzuki}]{sum07}
Sumiyoshi, K., Yamada, S., \& Suzuki, H. 2007, Astrophys.\ J., 667, 382

\bibitem[{Sumiyoshi {et~al.}(2005)Sumiyoshi, Yamada, Suzuki, Shen, Chiba, \&
  Toki}]{sum05}
Sumiyoshi, K., Yamada, S., Suzuki, H., {et~al.} 2005, Astrophys.\ J., 629, 922

\bibitem[{{Suwa}(2014)}]{suw14}
{Suwa}, Y. 2014, \pasj, 66, L1

\bibitem[{{Suwa} {et~al.}(2013){Suwa}, {Takiwaki}, {Kotake}, {Fischer},
  {Liebend{\"o}rfer}, \& {Sato}}]{suwa13}
{Suwa}, Y., {Takiwaki}, T., {Kotake}, K., {et~al.} 2013, \apj, 764, 99

\bibitem[{{Suwa} {et~al.}(2016){Suwa}, {Yamada}, {Takiwaki}, \&
  {Kotake}}]{suwa16}
{Suwa}, Y., {Yamada}, S., {Takiwaki}, T., \& {Kotake}, K. 2016, \apj, 816, 43

\bibitem[{{Suwa} {et~al.}(2018){Suwa}, {Yoshida}, {Shibata}, {Umeda}, \&
  {Takahashi}}]{suwa18}
{Suwa}, Y., {Yoshida}, T., {Shibata}, M., {Umeda}, H., \& {Takahashi}, K. 2018,
  \mnras, 481, 3305

\bibitem[{Suzuki(1993)}]{suz93}
Suzuki, H. 1993, in Proceedings of the International Symposium on Neutrino
  Astrophysics: Frontiers of Neutrino Astrophysics, ed. Y.~Suzuki \&
  K.~Nakamura (Tokyo: Universal Academy Press Inc.), 219

\bibitem[{Suzuki(1994)}]{suz94}
Suzuki, H. 1994, in Physics and Astrophysics of Neutrinos, ed. M.~Fukugita \&
  A.~Suzuki (Tokyo: Springer-Verlag), 763

\bibitem[{Suzuki(2005)}]{suz05}
Suzuki, H. 2005, in Proceedings of the Fifth International Workshop on Neutrino
  Oscillations and their Origin, ed. Y.~Suzuki, M.~Nakahata, S.~Moriyama, \&
  Y.~Koshio (Singapore: World Scientific), 332

\bibitem[{{Takiwaki} \& {Kotake}(2018)}]{tak18}
{Takiwaki}, T., \& {Kotake}, K. 2018, \mnras, 475, L91

\bibitem[{{Tamborra} {et~al.}(2013){Tamborra}, {Hanke}, {M{\"u}ller}, {Janka},
  \& {Raffelt}}]{tam13}
{Tamborra}, I., {Hanke}, F., {M{\"u}ller}, B., {Janka}, H.-T., \& {Raffelt}, G.
  2013, Physical Review Letters, 111, 121104

\bibitem[{{Thankful Cromartie} {et~al.}(2019){Thankful Cromartie}, {Fonseca},
  {Ransom}, {Demorest}, {Arzoumanian}, {Blumer}, {Brook}, {DeCesar}, {Dolch},
  {Ellis}, {Ferdman}, {Ferrara}, {Garver-Daniels}, {Gentile}, {Jones}, {Lam},
  {Lorimer}, {Lynch}, {McLaughlin}, {Ng}, {Nice}, {Pennucci}, {Spiewak},
  {Stairs}, {Stovall}, {Swiggum}, \& {Zhu}}]{tha19}
{Thankful Cromartie}, H., {Fonseca}, E., {Ransom}, S.~M., {et~al.} 2019, arXiv
  e-prints, arXiv:1904.06759

\bibitem[{{Thompson} {et~al.}(2003){Thompson}, {Burrows}, \& {Pinto}}]{tho03}
{Thompson}, T.~A., {Burrows}, A., \& {Pinto}, P.~A. 2003, \apj, 592, 434

\bibitem[{{Totani} {et~al.}(1998){Totani}, {Sato}, {Dalhed}, \&
  {Wilson}}]{tot98}
{Totani}, T., {Sato}, K., {Dalhed}, H.~E., \& {Wilson}, J.~R. 1998, Astrophys.\
  J., 496, 216

\bibitem[{{Umeda} {et~al.}(2012){Umeda}, {Yoshida}, \& {Takahashi}}]{umeda12}
{Umeda}, H., {Yoshida}, T., \& {Takahashi}, K. 2012, Progress of Theoretical
  and Experimental Physics, 2012, 01A302

\bibitem[{{Vogel} \& {Beacom}(1999)}]{1999PhRvD..60e3003V}
{Vogel}, P., \& {Beacom}, J.~F. 1999, \prd, 60, 053003

\bibitem[{{Woosley} \& {Weaver}(1995)}]{ww95}
{Woosley}, S.~E., \& {Weaver}, T.~A. 1995, \apjs, 101, 181

\bibitem[{{Yamada}(1997)}]{yama97}
{Yamada}, S. 1997, \apj, 475, 720

\bibitem[{{Yamada} {et~al.}(1999){Yamada}, {Janka}, \& {Suzuki}}]{yama99}
{Yamada}, S., {Janka}, H.-T., \& {Suzuki}, H. 1999, \aap, 344, 533

\bibitem[{{Yokozawa} {et~al.}(2015){Yokozawa}, {Asano}, {Kayano}, {Suwa},
  {Kanda}, {Koshio}, \& {Vagins}}]{yok15}
{Yokozawa}, T., {Asano}, M., {Kayano}, T., {et~al.} 2015, \apj, 811, 86

\bibitem[{Zhang {et~al.}(2016)}]{Super-Kamiokande:2015xra}
Zhang, Y., {et~al.} 2016, Phys. Rev., D93, 012004

\end{thebibliography}
\end{document}